\documentclass[a4paper,11pt]{article}

\usepackage{graphicx}
\usepackage{pslatex}
\graphicspath{{figures/}}
\newcommand{\z}[1]{\mathbf{#1}}

\begin{document}

\begin{center}
{\large Infinite non-causality in active cancellation of random noise}\\
{Emmanuel F{\textsc{riot}}}\\
{CNRS - Laboratoire de M\'ecanique et d'Acoustique,\\
$31$ chemin Joseph Aiguier, $13402$ Marseille, France}\\
e-mail: friot@lma.cnrs-mrs.fr\\
Tel.: +33 4 91 16 40 84\\
Fax: +33 4 91 16 40 80\\
\ \\
22 pages of text - 19 figures\\
\ \\
\end{center}

\begin{abstract}
Active cancellation of broadband random noise requires the detection of the incoming
noise with some time advance. In an duct for example this advance must be larger than
the delays in the secondary path from the control source to the error sensor.
In this paper it is shown that, in some cases, the advance required for perfect
noise cancellation is theoretically infinite because the inverse of the secondary path,
which is required for control, can include an infinite non-causal response.
This is shown to be the result of two mechanisms: in the single-channel case
(one control source and one error sensor),
this can arise because of strong echoes in the control path.
In the multi-channel case this can arise even in free field simply because of an unfortunate
placing of sensors and actuators.
In the present paper optimal feedforward control is derived through analytical and numerical
computations, in the time and frequency domains. It is shown that, in practice,
the advance required for significant noise attenuation can be much larger than the secondary path delays.
Practical rules are also suggested in order to prevent infinite non-causality from appearing.
\end{abstract}
 
\newpage
\section{Introduction}
\noindent

Active Noise Control (ANC) can help in reducing broadband random noises such as road noise
inside cars (see \textit{e.g.} Refs. $\lbrack$1--3$\rbrack$), fan noise travelling in ducts (Ref.~\cite{Roure})
or noise induced by the boundary-layer pressure fluctuations around aircraft fuselages (Refs.~$\lbrack$5--6$\rbrack$). 
However, in the broadband random case, the noise arriving at the minimization microphones
is mostly unpredictable, and the constraint of causality, which applies to the control filters,
can restrain the performances of ANC (cf.~Ref.~\cite{Nelson}).
Therefore, in practice, ANC set-ups often rely on feedforward control which
involves one or several reference sensors detecting the incoming noise
before it reaches the minimization microphones.
In the single-channel case of noise travelling along one direction, the constraint of causality
is easy to interpret in terms of acoustic propagation: the primary noise must
be detected before it reaches the minimization sensor, with a time advance larger than
the propagation time from the control source to the minimization microphone. The time advance
must also compensate the electrical and computation delays in the control processing unit.
However, in more complex cases, it is not easy to link the causality constraint to propagation
delays. Indeed, no design rule guarantees the causality of perfect noise cancellation
without effective computation of the optimal control filters.

In this context, this paper shows that perfect noise cancellation can require
control filters with an infinite non-causal response, which means that perfect
feedforward control can require the detection of the incoming noise with an infinite
time advance. Analytical computations on simple ANC set-ups show that
this infinite non-causality is the theoretical result of two mechanisms:
in the single-channel case of one control source and one minimization microphone,
it is the consequence of strong echoes in the secondary path.
In the multi-channel case, it can arise even in free field because of an unfortunate
geometrical arrangement of the actuators and sensors.
In the paper, the performance of causally-constrained ANC is also investigated
through numerical computations. It is shown that, in practice, achieving a significant
noise attenuation can require the detection of the incoming noise with a very large time advance.

The study presented in this paper was motivated by the unexpected difficulties encountered
when implementing multi-channel broadband ANC in a large room at LMA. 
Although many successful experiments of random noise control in enclosures had been reported
(see \textit{e.g.} Refs.~\cite{Joplin} and \cite{Laugesen}), it appeared
that efficient control in the LMA room required feeding the reference signal
to the control unit with a time advance of several seconds (see Ref.~\cite{Bordier}).
Previous papers have reported that the response of optimal control filters could include
a very long non-causal component (\textit{e.g.} in Refs.~\cite{Miyoshi}, \cite{Lopes} and \cite{Yuan}),
but, although a link between non-minimum phase responses and reverberation
had been observed earlier (see Ref.~\cite{Neely}), this has been seen in the 
context of ANC as a purely algebraic problem.
The present paper is an attempt to give an acoustic interpretation
of control non-causality in order to derive design guidelines for ANC set-ups.

In section~\ref{sec:1x1} of this paper, the causality of ANC is studied in the single-channel case.
After recalling the case of noise cancellation in an infinite duct, it is shown that
one echo in the secondary path can lead to infinitely non-causal control.
Optimal non-causal control and causally-constrained control are then derived for
the case of a single monopole source and a single pressure microphone in the corner
of two rigid walls.
Numerical simulations in the frequency and time domain show that, in some cases,
the incoming primary noise must be detected with a very large time advance for efficient control.
Finally it is shown that a slight change in the actuator and sensor locations can relax
the causality constraint, which suggests guidelines for the design of ANC set-ups
not prone to infinite non-causality.

In section~\ref{sec:2x2}, ANC causality is investigated in the multi-channel case.
Optimal non-causal control is derived analytically for a free-field set-up including
two monopole sources and two microphones, and some arrangements are shown to lead
to infinitely non-causal noise cancellation. The performance of causally-constrained control
is evaluated through numerical simulations involving a Filtered-Reference Least Mean Square
(FXLMS) algorithm. In this 2x2 case it is shown that non-causality is critical when
the condition number of the matrix of secondary paths is high at low-frequency.

The inversion of acoustic paths, which is required for real-time ANC, is also needed 
in devices for "virtual acoustics" which aim at a sound reproduction conveying
accurate information such as the location of the recorded noise sources (see Ref.~\cite{Nelson-Kirkeby}).
Since sound reproduction is not a real-time process, the non-causality
arising from the acoustic path inversion is usually not critical.
However this paper shows that care must be taken for this inversion,
even in the context of virtual acoustics, because the additional advance required
for causal inversion can be much larger than the delays in the direct path.

\section{Non-causality in the single-channel case with reflections} \label{sec:1x1}
\subsection{Feedforward control without reflections} \label{sec:duct}
\noindent
Figure~\ref{fig:feedforward_set-up} displays a standard feedforward set-up for cancellation
of random noise travelling at low-frequency in a duct (see Ref.~\cite{Nelson}). The duct is assumed to
be rigid and either infinite or with anechoic terminations. A directional microphone detects the
incoming noise (and not the secondary noise from the control source).
An adaptive controller $W$ filters this reference signal to generate appropriate input $u$
at the control source to minimize noise at the error sensor.
Below the first cut-off frequency of the duct, the noise $e$ at the error sensor can be
ideally written in the frequency domain as the sum of the primary and secondary noises:
\begin{equation}
e = e^{-j\omega L /c} x + \frac{\rho q c}{2S} e^{-j\omega l /c} = F x + H q
\end{equation}
where $c$ is the speed of sound, $\rho$ the air density, $S$ the cross-sectional area of the duct
and $q$ the loudspeaker volume velocity. If the loudspeaker dynamics is neglected so that
$q$ is directly the output of the control filter, the noise $e$ at the error sensor will be cancelled
if:
\begin{equation}
q=Wx=-FH^{-1}x=-\frac{2S}{\rho c}e^{-j\omega (L-l) /c}x
\end{equation}
It appears that the volume velocity $q$ which cancels noise is the
causal filtering of the signal $x$ if the reference sensor is located
\emph{upstream} of the loudspeaker (i.e. $L\geq l$).
In practice however the secondary path $H$, which has to be inverted for perfect control,
includes the dynamics of the loudspeaker and the delays inherent to anti-aliasing
and reconstruction filters in a digital adaptive controller. The detection sensor
must therefore be located further upstream in order to provide a reference signal
with an advance larger than the \emph{overall} delay in the secondary path.
The purpose of this paper is to show that this classical result, which can be found in the textbooks on ANC
(cf. Ref.~\cite{Nelson}), does not apply to single-channel cases involving noise reflections
nor to multi-channel cases even in free field: sometimes the detection sensor must provide
a reference signal with an infinite advance for perfect control.

\subsection{Feedforward control with one reflection} \label{one reflection}
\noindent

In this section it is supposed that the transfer function $H$ between the control source and
the error signal is the sum of direct propagation, with delay $\tau_1$ and arbitrary magnitude $1$,
and of one echo with delay $\tau_2$ ($\tau_2>\tau_1$) and complex magnitude $\alpha$ resulting
from one (possibly non-rigid) acoustic reflexion or from the combination of several reflexions
arriving simultaneously at the error sensor : 
\begin{equation}
H(\omega)=e^{-j\omega \tau_1}+\alpha e^{-j\omega \tau_2}
\end{equation}
If the direct sound is louder than the echo ($|\alpha|<1$), the inverse of the transfer
function $H^{-1}$ can be expanded as:
\begin{equation} \label{causal}
H^{-1}  =  \frac{1}{e^{-j\omega \tau_1} + \alpha  e^{-j\omega \tau_2}}
= e^{j\omega \tau_1} \left[ 1 - {\alpha} e^{-j\omega (\tau_2-\tau_1)}
	+ {\alpha^2} e^{-2j\omega (\tau_2-\tau_1)}-\ldots \right]
\end{equation}
Therefore, in the time domain, $H^{-1}$ has an Infinite Impulse Response including a non-causal
part with finite length $\tau_1$. As in the infinite duct case, the optimal control filter $-FH^{-1}$
will be fully causal if the transfer function $F$ from the reference sensor to the error sensor
includes a delay larger than the propagation time from the control source to the error sensor.
When the primary noise is a pulse, the infinite time length of $H^{-1}$ can be interpreted in a similar way
to that applicable to cross-talk cancellation (cf. Ref.~\cite{Nelson-Kirkeby}): the expansion in (\ref{causal})
shows that the primary noise is cancelled by the direct sound resulting from a pulse in the control source,
and that the echo of the control pulse is in turn cancelled by the direct sound of a second control pulse;
and so on (the echo of the $n^{\mathrm{th}}$ control pulse is cancelled by the direct sound of a $(n+1)^{\mathrm{th}}$
control pulse). Since the direct sound is louder than its echo, an infinite but converging series
of pulses is required to achieve perfect cancellation.

However the echo can be \emph{louder} than the direct sound ($|\alpha|>1$).
This may seem unlikely at first sight but a simple acoustic set-up with this feature is
shown in section \ref{sec:corner}. The inverse transfer function $H^{-1}$ can then be
expanded as:
\begin{equation} \label{eqn:serie}
H^{-1}  = \frac{\frac{1}{\alpha} e^{j\omega \tau_2}}{1 + \frac{1}{\alpha}  e^{-j\omega (\tau_1-\tau_2)}} 
=  \frac{1}{\alpha} e^{j\omega \tau_2} \left[ 1 - \frac{1}{\alpha} e^{j\omega (\tau_2-\tau_1)}
	+ \frac{1}{\alpha^2} e^{2j\omega (\tau_2-\tau_1)}-\ldots \right]
\end{equation}
In this case it appears that $H^{-1}$ has an infinite non-causal response,
such that causal noise cancellation requires
detection of the primary noise with an infinite time advance.
With a pulse as primary noise, the theoretical non-causality of control can be interpreted
from the expansion in (\ref{eqn:serie}): the primary noise is cancelled by the echo
of a counter-pulse in the control source, and the direct sound of this counter-pulse
is pre-cancelled by the echo of a previous pulse, and so on. In the end an infinite converging series
of control pulses in the past is required for noise cancellation of a single primary pulse.

The simple analysis above shows that ANC is infinitely non-causal as soon as the echo
is louder than the direct sound in the response.
However equation (\ref{eqn:serie}) also demonstrates that, if the echo is \emph{very much} louder,
coefficient $\alpha$ is large and the series in (\ref{eqn:serie}) converges quickly.
In this case the non-causal terms for large negative times may be neglected. Eventually control
may be critically non-causal especially when the echo is slightly louder than the direct sound.

\subsection{Optimal non-causal control in a corner} \label{sec:corner}
\noindent

Figure \ref{fig:coinJSV} shows an ANC set-up where the secondary path
includes an echo louder than the direct sound: two infinite rigid planes
define in 3D a corner with a directional detection sensor, an omnidirectional
error sensor and a control source in the bisecting plane. With the notations
from figure \ref{fig:coinJSV} and if the control source can be seen
as a 3D monopole, which is the case of a loudspeaker at low frequency,
the secondary transfer function $H$ from the loudspeaker volume acceleration
$\dot{q}$ to the error signal $e$ can be computed by taking image sources into account:
\begin{equation}
 H(\omega) = \frac{\rho}{4\pi} \left[ \frac{e^{-j\omega L_1/c}}{L_1} 
+ 2 \frac{e^{-j\omega L_2/c}}{L_2}  + \frac{e^{-j\omega L_3/c}}{L_3} \right]
\end{equation}
with $L_1=L$, $L_2^2=(L_1+l)^2+l^2$ and $L_3=L_1+2l$.
If $L_2$ is less than $2L_1$ then the first echo in the secondary path is louder than
the direct path. It can be shown that this condition amounts to
 $L>\frac{1+\sqrt{7}}{3}l$. Figure \ref{fig:second_path_freq} displays the secondary path
as a function of frequency when $c=340m/$s, $L=1$m and $l=0.8$m (in fact these distances
have been slightly adjusted in the subsequent numerical simulations so that noise propagation
over $L$ and $l$ takes an integer number of samples). Figure \ref{fig:second_path_time}
displays the corresponding exact Impulse Response and its digital approximation
computed by Inverse Fast Fourier Transform for a sampling frequency of $16384Hz$ and
a frequency resolution of $\frac{1}{60}Hz$.

When the incoming primary noise is a plane wave propagating along the bisecting plane,
as sketched in figure \ref{fig:coinJSV}, the transfer function $F$ from the directional
reference sensor (detecting the incident plane wave only) to the
omnidirectional error sensor (detecting the incident \emph{and} reflected waves) is given
for the primary noise by:
\begin{equation}
F(\omega)= \frac{e^{-j\omega(L+l)/c}+e^{+j\omega(L+l)/c}}{e^{j\omega(2L+l+\Delta)/c}}
\end{equation}
In this equation $\Delta$ denotes the distance between the reference sensor and the
error sensor, as shown in figure \ref{fig:coinJSV}.  
Without taking the dynamics of a real loudspeaker into account, perfect noise cancellation
at the error sensor will be achieved if the input to the control source is equal to the reference
signal filtered by the optimal filter $W$:
\begin{equation}
W = - H^{-1}F = - \frac{4\pi}{\rho} \frac{e^{-j\omega(L+\Delta)/c}+ e^{-j\omega(3L+2l+\Delta)/c}}
{\frac{e^{-j\omega L_1/c}}{L_1} + \frac{2e^{-j\omega L_2/c}}{L_2}  + \frac{e^{-j\omega L_3/c}}{L_3}}
\end{equation}
With two echoes in the secondary path, no series expansion similar to that of equation (\ref{eqn:serie})
can easily express the causality of the optimal control filter $W$. However the Impulse
Response of $W$ can be computed numerically by IFFT. Figure~\ref{fig:opt_temp} displays
the Impulse Response of $W$ when $\Delta=0$m for a sampling frequency of $16384Hz$
and a frequency resolution of $\frac{1}{120}Hz$. In the case without reflections, $\Delta\geq 0$
would have enforced the control causality but, in the corner, the response includes a very long non-causal component,
with significant coefficients at negative times much larger than any propagation time between
the elements of the set-up in figure \ref{fig:coinJSV}. This means that noise cancellation could
be achieved only by moving the detection sensor far away from the corner (i.e. by enforcing
$\Delta>>0$), which may not be possible in practice (e.g. if the primary noise source is not far away).

\subsection{Optimal causally-constrained control in a corner} \label{sec:optcaus}
\noindent

In the previous section it has been shown that \emph{perfect} noise cancellation can require
very long non-causal filters. In practice however, the aim of ANC is only to provide 
\emph{significant} noise attenuation. To this intent
the performance of causal control must be determined as a function of the advance
with which the reference signal is provided to the feedforward controller.

As a first try, the quantitative effect of the causality constraint can be assessed by
computing the noise attenuation achieved with the causal control filter obtained
discarding the non-causal coefficients in the time response of the optimal filter.
For the set-up shown in figure \ref{fig:coinJSV} with $L=1$m and $l=0.8$m,
figure \ref{fig:atten_tronq} displays the noise attenuation achieved with this causally-truncated
filter, when $\Delta=0$m, $\Delta=136$m and $\Delta=1700$m, which amounts to providing
the reference signal with an advance of respectively $0$s, $0.4$s and $5$s.
Figure \ref{fig:atten_tronq} shows that a huge advance in the reference signal 
is required to achieve substantial active noise reduction with a causally-truncated control filter.

However the causal truncation of the non-causal optimal control filter is \emph{not} the optimal
causally-constrained control filter. The causality constraint does not restrain the control
performance as much as figure \ref{fig:atten_tronq} suggests.
The optimal causal filter depends on the predictability of the primary noise. Following Ref.~\cite{Elliott},
it can be written in the case of white noise as:
\begin{equation} \label{eq:opt-caus}
W = - \frac{1}{H_{min}} \left\{ \frac{F}{H_{all}} \right\}_+ 
\end{equation}
where $H=H_{min}H_{all}$ is the factorization of the secondary path FRF $H$ into minimum-phase
and all-pass components $H_{min}$ and $H_{all}$, and $\{\}_+$ means that
the non-causal component of the quantity enclosed by the braces is
discarded. As in the previous section, $F$ denotes the transfer function from the detection
sensor to the error sensor for the primary noise and $H$ the secondary path transfer function.

Figure \ref{fig:atten_opt_caus} displays the noise attenuation achieved with the causally-constrained
optimal control filter when the reference signal is provided with an advance of $0$s,
$\frac{512}{16384}=0.03125$s and $0.4$s, which amounts respectively
to $\Delta=0$m, $\Delta=10.625$m and $\Delta=136$m. 
For this figure the factorization of $H$ into minimum-phase
and all-pass components was performed in the frequency domain with the complex-cepstrum method (cf. Ref.~\cite{Oppenheim}).
For a given advance in the reference, the optimal causal filter performs much better than the
causally truncated non-causal filters: with a $0.4$s advance, optimal causal control leads to a large noise reduction
as in the case of causally-truncated non-causal control with a $5$s advance.
However the reference signal must still be provided with a very large advance for efficient control.
Figure \ref{fig:opt_caus_temp} displays the Impulse Response of the causal filter for
$\Delta=0$m and $\Delta=136$m. For $\Delta=0$m it shows that the optimal causal filter
is very different from the causal truncation of the optimal non-causal controller, but for
$\Delta=136$m it confirms that, with a very large advance in the reference, the response
is close to the time-shifted response of the optimal non-causal controller
(displayed in figure \ref{fig:opt_temp}).

In practice ANC implementations very often rely on adaptive transverse
control filters whose coefficients are optimized recursively in the time-domain
with the Filtered-reference(X) Least-Mean Square algorithm (cf. Ref.~\cite{Nelson}).
Figure \ref{fig:atten_fxlms_aver} displays the noise attenuation achieved
with a 1024-tap control filter continuously updated with an FXLMS algorithm
for $\Delta=0$m, $\Delta=10.625$m and $\Delta=21.25$m, and primary white
noise. The propagation time over $\Delta=21.25$m corresponds to the whole length of
the control filter impulse response.
The FXLMS converged
very slowly for these results. Even after selecting the best convergence coefficient,
several seconds were typically required before achieving
noise attenuation, and leakage (cf. Ref.~\cite{Elliott}) did not improve convergence.
For $\Delta=10.625$m the attenuation is of the same order of magnitude as for the optimal
causal filter computed in the frequency domain.
The zeroes in the attenuation are at the same frequency as the zeroes of the secondary path
$H$ or of the reference signal spectrum $|F|^2$. At these frequencies the FXLMS converges
with difficulties (cf. Ref.~\cite{Elliott}).
It could be expected that detecting the reference signal at a distance larger
than $\Delta=10.625$m improve control performance, as for the
causally-constrained filters derived in the frequency-domain. However the poor
result shown in figure \ref{fig:atten_fxlms_aver} for $\Delta=21.25$m means
that the optimum advance of the reference signal does depend on the length
of the filter response in the case of FXLMS control; the best control
performance is not achieved with the largest pre-delay when the control filter
impulse response is too short. 

Figure \ref{fig:fxlms_temp} shows the impulse response of a FIR control filter
obtained when the adaptation process is frozen after minimization of the error signal.
Figure \ref{fig:atten_fxlms_froz} displays the noise attenuation theoretically
achieved with this specific control filter. The computation is performed in the
frequency domain for purely white primary noise.
The attenuation for $\Delta=10.625$m is clearly worse than the attenuation displayed
in figure \ref{fig:atten_fxlms_aver} for a continuously adapting control filter.
Therefore the permanent adaptation of the control filter helps in ensuring good performance
of FXLMS control, even after the residual error has apparently converged;
a linear control FIR filter with permanent coefficient adaptation is in fact a non-linear
control filter which sometimes performs much better than any stationary
linear control filter (see Ref.~\cite{Haykin}).

\subsection{Precautions against non-causality} \label{precautions}

The series expansion in equation~(\ref{eqn:serie}) suggests that control is
critically non-causal when the echo in the secondary path response is only slightly
louder than the direct sound. For the set-up of figure \ref{fig:coinJSV} the first echo
is as loud as the direct sound if $L=\frac{1+\sqrt{7}}{3}l\approx 1.2153l$. With
$l=0.8$m this limit amounts to $L\approx0.9722$m, which is close to the value $L=1$m chosen
for the numerical simulations. 

Figure \ref{fig:coin_moins_loin} displays the impulse response of the non-causal
optimal filter $W=-H^{-1}F$ for $L=0.95$m. The control filter has a very
long response with only a relatively short (but non-zero) non-causal component.
Figure \ref{fig:coin_plus_loin} displays the impulse response for $L=1.05$m. In this case
the non-causal response decays more rapidly than for the case $L=1$m.
These figures show that a small change in length $L$ can dramatically
affect the causality of optimal control. Control is highly non-causal only for the
values of $L$ which lead to a first echo slightly louder than the direct sound.

In practice ANC set-ups often involve secondary paths with complicated time response
including many echoes. It is not easy to detect without measurements if a given set-up
gives rise to non-causality because of reflections, even if, as the previous simulations
suggest, this will be critical only for very specific arrangements.
One simple step towards preventing infinite non-causality in the single-channel
case with reflections is to move the control source and the error sensor as close as possible to
each other.
In this way the direct sound is maximized. Furthermore
reducing delays in the secondary path improves the convergence of adaptive algorithms
such as the FXLMS (cf. Ref.~\cite{Elliott}). However in practice other constraints may
impose a minimum distance in the secondary path. For example the error sensor must
usually be in the far-field of the control source to achieve noise reduction over
more than a very narrow area. 
In the end infinite non-causality must be remembered as a theoretical possibility
in active control of broadband noise.

\section{Non-causality in the multi-channel case in free field} \label{sec:2x2}
\subsection{Optimal Control in a 2x2 case}
\noindent
Figure~\ref{fig:2-2set-upJSV} shows a set-up for ANC in free field with two secondary sources
and two omnidirectional minimization microphones. The incoming primary noise is assumed to be a plane wave.
The loudspeakers are idealized as monopoles with volume accelerations $\dot{q}_1$ and $\dot{q}_2$.
The microphone \#1 is taken as the reference sensor for feedforward control. The feedback from the control
sources to the reference signal $x$ will not be considered in the derivation of optimal control. In practice
this can be done if the controller has an Internal Model Control structure
including cancellation of the feedback path (see Ref.~\cite{Morari}). It is also assumed that the output control
filters $W_1$, $W_2$ with input $x$ are directly the volume accelerations $\dot{q}_1$, $\dot{q}_2$ of
the control sources, the dynamics of the loudspeakers is not taken into account.

With the notations from figure~\ref{fig:2-2set-upJSV} and with $k=\omega/c$, the total
noises $p_1$, $p_2$ at the minimization microphones can be written as:
\begin{equation}
\left[ 	\begin{array}{c} p_1\\p_2 \end{array}
\right] = \; \left[ 	\begin{array}{c} 1\\ e^{-jkd_5\cos\theta}\end{array} \right] x
+ \frac{\rho_0}{4\pi}\;\ \left[ 
	\begin{array}{cc} 
		\frac{e^{-jkd_1}}{d_1} & \frac{e^{-jkd_2}}{d_2}\\
		\frac{e^{-jkd_3}}{d_3} & \frac{e^{-jkd_4}}{d_4} 
	\end{array}
\right]
\left[ 	\begin{array}{c} W_1\\W_2 \end{array} \right] x
\end{equation}
Noise can be cancelled at both error sensors and at all frequencies
if $d_1d_4 \neq {d_2d_3}$. The optimum control filters $W_1$, $W_2$ are then given by:
\begin{equation}\label{eqn:optimal}
\left[ 	\begin{array}{c} W_1\\W_2 \end{array}
\right]
 = -\;\frac{4\pi}{\rho_0}\;\frac{d_1 d_2 d_3 d_4}{d_2d_3 e^{-jk(d1+d4)}-d_1d_4e^{-jk(d2+d3)}}
 \left[ 
	\begin{array}{cc} 
		\frac{e^{-jkd_4}}{d_4} & -\frac{e^{-jkd_2}}{d_2}\\
		-\frac{e^{-jkd_3}}{d_3} & \frac{e^{-jkd_1}}{d_1} 
	\end{array}
\right]
\left[ 	\begin{array}{c} 1\\e^{-jkd_5\cos\theta} \end{array} \right]
\end{equation}
It can be assumed, without loss of generality because the
indexes can be switched, that $d_1d_4 < d_2d_3$. After introducing
$\alpha=d_1d_4/d_2d_3$ and $D=d_2+d_3-d_1-d_4$, equation (\ref{eqn:optimal}) can be
re-written and expanded as:
\begin{eqnarray} \label{eqn:optimal3}
\left[ 	\begin{array}{c} W_1\\W_2 \end{array}
\right] = -\;\frac{4\pi d_1 d_4}{\rho_0} && \!\!\!\!\!\!\left( 1 + \alpha e^{-jkD}
-\alpha^2 e^{-j2kD} + \ldots\right) \nonumber \\
 &&\left[ 
	\begin{array}{cc} 
		\frac{e^{jkd_1}}{d_4} & -\frac{e^{jk(d_1+d_4-d_2)}}{d_2}\\
		-\frac{e^{jk(d_1+d_4-d_3)}}{d_3} & \frac{e^{jkd_4}}{d_1} 
	\end{array}
\right]
\left[ 	\begin{array}{c} 1\\e^{-jkd_5\cos\theta} \end{array} \right]
\end{eqnarray} 
The causality of perfect noise cancellation can be discussed from this expression.
Firstly, $e^{-jkd_5\cos\theta}$ is a time delay if $\cos\theta>0$, that is if the
primary noise is impinging the reference microphone before all the minimization microphones.
In the opposite case $e^{-jkd_5\cos\theta}$ denotes a finite time advance equal to the propagation time
of the primary noise between the reference microphone and the minimization microphones
which receives it first.
Secondly, the matrix in equation (\ref{eqn:optimal3}) includes finite time advances.
If $D>0$ one has $d_1+d_4-d_2<d_3$ and $d_1+d_4-d_3<d_2$, 
which means that these time advances are always smaller than the largest
propagation time in the secondary path. In this case where $D>0$,
the series expansion in equation (\ref{eqn:optimal3}) will also be fully causal. Therefore,
if $D>0$, the single-channel result without reflections of section \ref{sec:duct} generalizes:
noise cancellation is causal if the incoming primary noise is detected before it
impinges the first minimization microphone, with a time advance larger than the largest
propagation time in the secondary path.
However, if $D<0$, the series expansion in equation (\ref{eqn:optimal3}) gives rise to
an infinitely-long non-causal impulse response for optimal control. This can occur indeed,
for example in the case $d_1=2.8$m, $d_4=8.8$m, and $d_2=d_3=5$m.
By inspecting the case $d_1d_4> d_2d_3$, it finally appears that noise cancellation will be
infinitely non-causal in the general case if:
\begin{equation}\label{criterion}
(d_1d_4-d_2d_3)(d_1+d_4-d_2-d_3)<0
\end{equation}

Figure~\ref{fig:4cases} displays four arrangements with increasing distances $d_1$ and
$d_4$. At first sight it is not easy to detect significant differences between the
arrangements but, according to equation~(\ref{criterion}), the arrangements of figure~\ref{fig:4cases}(a)
and \ref{fig:4cases}(b) lead to infinitely non-causal noise cancellation.
Figure \ref{fig:rep4cases} displays the Impulse Responses of the optimal control filters,
computed by IFFT with a sampling frequency of 16384Hz, when the primary noise is
a plane wave with incidence angle $\theta=0^{\mathrm{o}}$.
The response for case~\ref{fig:4cases}(a) is non-causal but it decays
rapidly for negative times. Case~\ref{fig:4cases}(b) leads to a very long non-causal response,
which illustrates the non-causality of the series in equation~\ref{eqn:optimal3}.
The response for case~\ref{fig:4cases}(c) has a finite short-length non-causal component but
the causal component is very long, which would be a problem for practical implementation, whereas
the response for case~\ref{fig:4cases}(d) should pose no problem in practice.

The criterion given in equation~\ref{criterion} can be interpreted by considering
active control in the 2x2 case as the combination of
control with two individual single-channel devices and of cross-talk cancellation.
Firstly, parameter $\alpha={d_1d_4}/{d_2d_3}$ can be seen as the averaged magnitude of the cross-talk
at the minimization sensors. It quantifies the average ratio between the
"direct" sound received by each of the sensor by the corresponding single-channel
source and the "cross-talk" sound coming from the other source;
if the arrangement of sensors and actuators is fully symmetric (\textit{i.e.}
$d_1=d_4$ and $d_2=d_3$),
$\alpha$ exactly measures this ratio at the two sensors. Secondly, parameter
${D}/{c}=(d_2+d_3-d_1-d_4)/c$ measures the averaged difference between the times of arrival
of the "direct" and "cross-talk" sounds at the minimization microphones.
If $\alpha<1$ and $D>0$, in average the "direct" sound is louder than the cross-talk
and it reaches the minimization microphone before.
If $\alpha<1$ and $D<0$, the cross-talk is not louder
but it has to be cancelled before the direct sound, which requires an infinitely long non-causal
response. If $\alpha>1$, the indexes of the sources can be switched for interpretation
but, alternatively, the cross-talk can be seen as an echo louder than the direct
sound if $D>0$, and the discussion of section \ref{one reflection} on control of
an echo louder than the direct sound applies. Although there is no acoustic reflection
in the free-field set-up of figure~\ref{fig:2-2set-upJSV}, the cross-talk can eventually be
seen as an echo whose control implies non-causality if it is louder than the direct noise,
as in the single-channel case with reflections.

\subsection{Causally-constrained control in a 2x2 case}

\noindent
Figure~\ref{fig:atten_tronq2x2} displays the noise attenuation achieved with causally-truncated
control for the arrangement of figure~\ref{fig:4cases}(b).
For these simulations in the frequency domain, the reference signal was fed to the controller
with the additional time advances $\tau=d_1/c=8.24$ms, $\tau=1$s and $\tau=2$s.
Figure~\ref{fig:atten_tronq2x2} shows that very large advances are required
for large attenuations. The finite advance $\tau=d_1/c$ is the minimum advance required for 
the secondary noise to reach all the minimization microphones before the primary noise;
it compensates the finite time advances which appear in equation~(\ref{eqn:optimal3}).
If optimal control were not infinitely non-causal, this finite time advance would be
large enough for optimal control to be perfectly causal, and an infinite noise attenuation
should be achieved. The very poor attenuation shown in figure~\ref{fig:atten_tronq2x2} for
$\tau=d_1/c$ illustrates the dramatic effect of non-causality upon control performances.
In figure~\ref{fig:atten_tronq2x2} the noise attenuation is higher at minimization microphone~\# 2
because of the propagation time of the primary noise between the two microphones, which is extra time
for causal control.

In the multi-channel stationary case, the optimal causally-constrained control filter can be
written as in equation~(\ref{eq:opt-caus}) as a function of the minimum-phase and all-pass
components of the matrix of secondary paths~\cite{Elliott}.
However, spectral factorization, which is an intermediate step in computing the optimal controller,
is considerably more difficult to perform numerically
in the multi-channel case than in the single-channel case (cf.~\cite{Sayed}). 
Instead, in the context of active control, it is more efficient to compute
sub-optimal controllers using adaptive algorithms in the time-domain.
Figure~\ref{fig:atten_fxlms_aver2x2} displays the noise attenuation achieved
with a 1024-tap control filter continuously updated with an FXLMS algorithm,
for several additional advances $\tau$ in the reference signal.
The FXLMS took several seconds to converge to the results shown in
figure~\ref{fig:atten_fxlms_aver2x2}. For $\tau=8.24$ms, control
performs poorly and the noise is not reduced at all at microphone~\#1. For $\tau=31.25$ms some noise
reduction is achieved at the two minimization microphones. However, in this case, the noise reduction
is low at some frequencies; it can be checked that the matrix of secondary paths is poorly conditioned
at these frequencies, which is known as slowing down the convergence of the FXLMS algorithm
(see~Ref.~\cite{Elliott}). Finally, the noise reduction is poor again for $\tau=62.5$ms.
As in section~\ref{sec:optcaus} for the single-channel case, this shows that,
when optimal control is largely non-causal, the performance of Active Noise Control
using a relatively short-length FIR filter depends strongly on the timing with which the incoming noise is
detected.

\subsection{Precautions against non-causality}
\noindent

In section \ref{precautions} it was shown that, in the single-channel case, non-causality
was especially critical for arrangements where a slight change in the geometry would
change the control filter with a non-causal response into a control filter with a long
causal response. For the multichannel case of figure~\ref{fig:2-2set-upJSV}, the series
in equation (\ref{eqn:optimal3}) shows that the filter Impulse Response is long when
$\alpha\approx 1$, which means that the secondary path $\z{H}$ is close to a singular matrix
at all frequencies. Therefore, in more complicated cases, the condition number of the
secondary path matrix at low-frequency may be an indicator of possible non-causality.

Figure~\ref{fig:conditionnement} displays the condition number at low-frequency
(\textit{i.e.} for $k\to 0$) of the secondary path matrix $\z{H}$, as a function of
parameter $d_1$ when $d_4=d_1+6$m and $d_2=d_3=5$m. The vertical lines in the plot
delimit the values of $d_1$ for which control is infinitely non-causal according
to equation~(\ref{criterion}). 
The four cases shown in figures~\ref{fig:4cases} and \ref{fig:rep4cases} are marked out
by stars in figure~\ref{fig:conditionnement}.
It appears that very long non-causal Impulse Responses were obtained for the
control filters when the condition number of the secondary matrix was high.
In this 2x2 case, good conditioning appears to be a protection from the critical
non-causality arising from the inversion of the secondary path matrix.
Furthermore, for adaptive feedforward control, good conditioning of the secondary path is
required for efficient convergence of the FXLMS algorithm (see Ref.~\cite{Elliott}).
In section \ref{precautions} a connection was also made in the single-channel case
with reflections between causality and FXLMS convergence. From these two simple
cases it may be inferred in practice that poor convergence of the FXLMS algorithm
may be the sign of optimal control non-causality.

As in the single-channel case, non-causality can be prevented in the case of
figure~\ref{fig:2-2set-upJSV} by bringing the loudspeakers and the error microphones
in closer pairs. In this way the condition number of the secondary path matrix
is reduced because the diagonal of the matrix is made dominant. Regularization techniques
can also be used to dampen the effect of poor matrix conditioning, e.g. by including
a leakage coefficient in the FXLMS adaptation formula (see Ref.\cite{Elliott}).
Finally, the author believes that improving the condition number of the matrix of
secondary paths also helps in preventing control non-causality in multi-channel
set-ups involving more than two actuators and sensors.
However, although non-causal set-ups with more channels can be built by adding
sensor-actuator pairs to the set-up of figure~\ref{fig:4cases}(b),
a connection between causality and matrix condition number could not be made
in the general multi-channel case.

\section{Conclusion}
\noindent

In this paper it has been shown that active cancellation of broadband noise
can theoretically require control
filters that include an infinitely non-causal impulse response. In the single-channel case,
this can happen because of high-level sound reflections. In the multi-channel case, 
the analysis of a 2x2 set-up has shown that this can occur
even in free field because of the algebraic inversion of the secondary path matrix
required for perfect cancellation. The cross-talk cancellation in the multi-channel
case can be seen as equivalent to the cancellation of primary noise echoes in the single-channel case.
Numerical simulations in the time and frequency domains have shown that, in practice,
situations are possible where a significant noise reduction cannot be achieved
without detecting the incoming primary noise with an advance very much larger
than the propagation time delays in the secondary path.

The inverse of the secondary path, which can have an infinite non-causal response,
is required not only for Active Noise Cancellation but also for accurate reproduction
of sound fields in applications such as "virtual acoustic imaging" (see
Ref.~\cite{Nelson-Kirkeby}). In these applications, the computation of the inverse path
need not be performed in real time, as is the case for ANC, and
a processing time can be afforded which reduces the consequences of non-causality.
However, designers of virtual acoustics devices must be aware 
that inverse acoustic responses can include time advances very much larger than
the propagation times or the reverberation time in the direct response.

It is not easy to prevent a large non-causality of the inverse acoustic responses
in practical complicated cases, especially in rooms with many resonant modes.
The analysis of the single-channel case in a corner and of the 2x2 case in free field
have shown that one possibility is to mount actuators and sensors in close pairs, so that sound
reflections and cross-talk components have lower levels than the direct sounds. This approach also
improves the rate of convergence of adaptive control algorithms such as the
Filtered-Reference Least Mean Squares. It also fits in with the current
trend of designing ANC devices which are decentralized, in order to
reduce the complexity of systems with many channels.
However the sensors must usually be in the far field of the actuators for global
active control. Therefore infinite non-causality cannot always be prevented,
and must be kept in mind as an ever possible limitation for active control of
random noise.

\newpage
\begin{center}
{\large List of figure captions}
\end{center}

\begin{figure}[h!]
\vspace*{5cm}
\centering
\includegraphics[width=12cm]{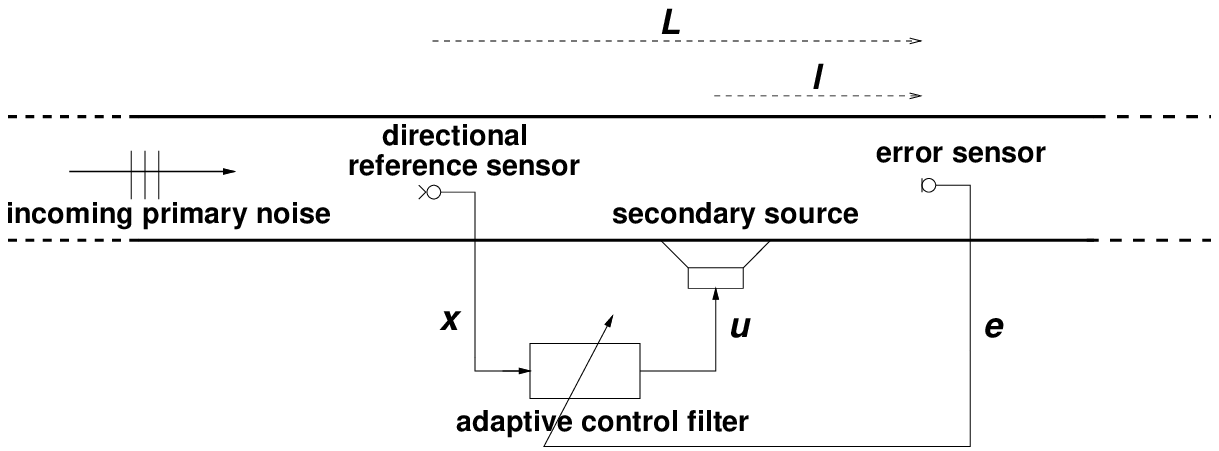}
\vspace*{5cm}
\caption{}
\label{fig:feedforward_set-up}
\end{figure}

Figure \ref{fig:feedforward_set-up}: A typical feedforward Active Noise Control set-up without reflections

\begin{figure}[ht]
\vspace*{3cm}
\centering
\includegraphics[width=10cm]{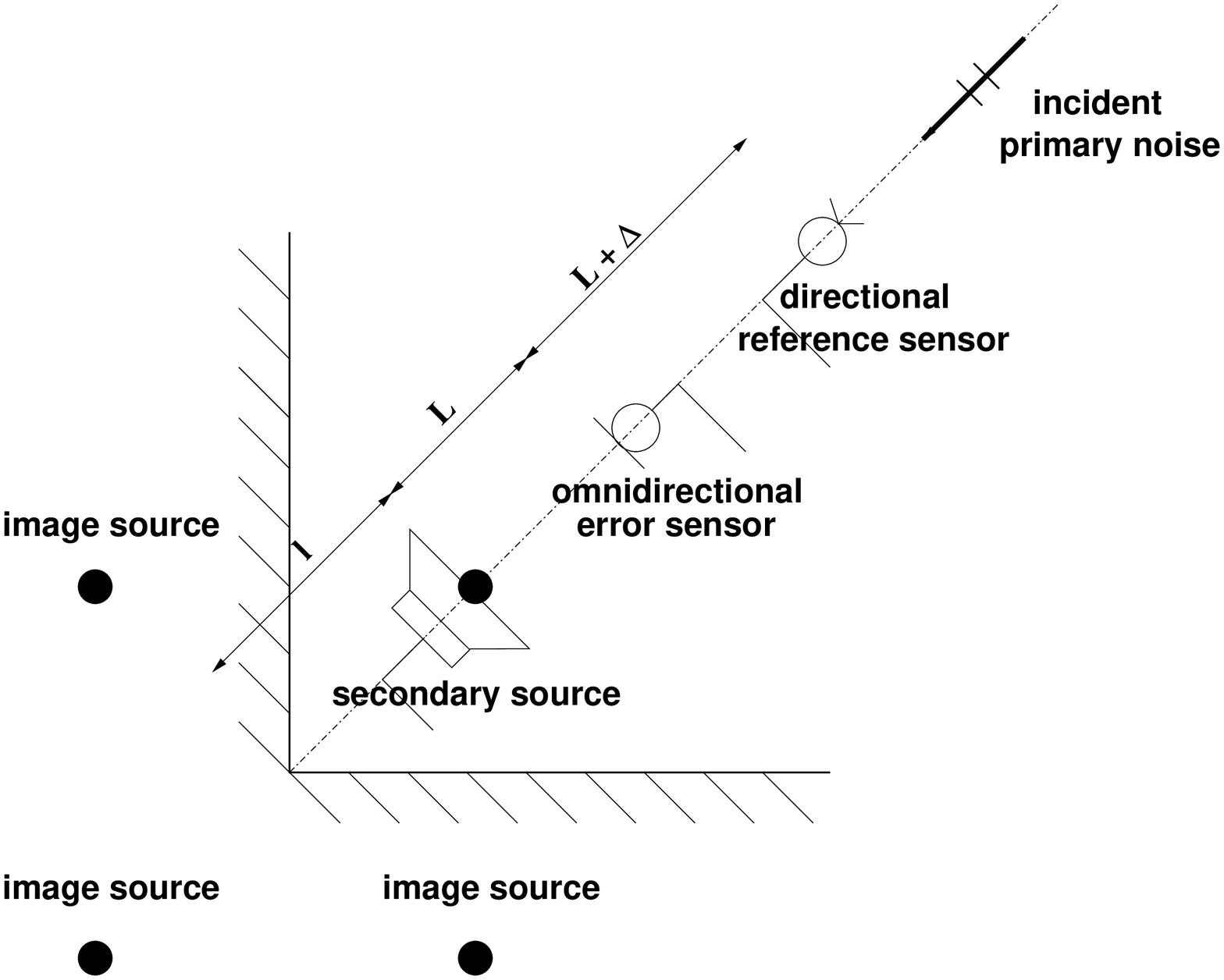}
\vspace*{5cm}
\caption{}
\label{fig:coinJSV}
\end{figure}

Figure \ref{fig:coinJSV}: A feedforward Active Noise Control set-up in a corner.

\begin{figure}[ht]
\vspace*{3cm}
\centering
\includegraphics[width=10cm]{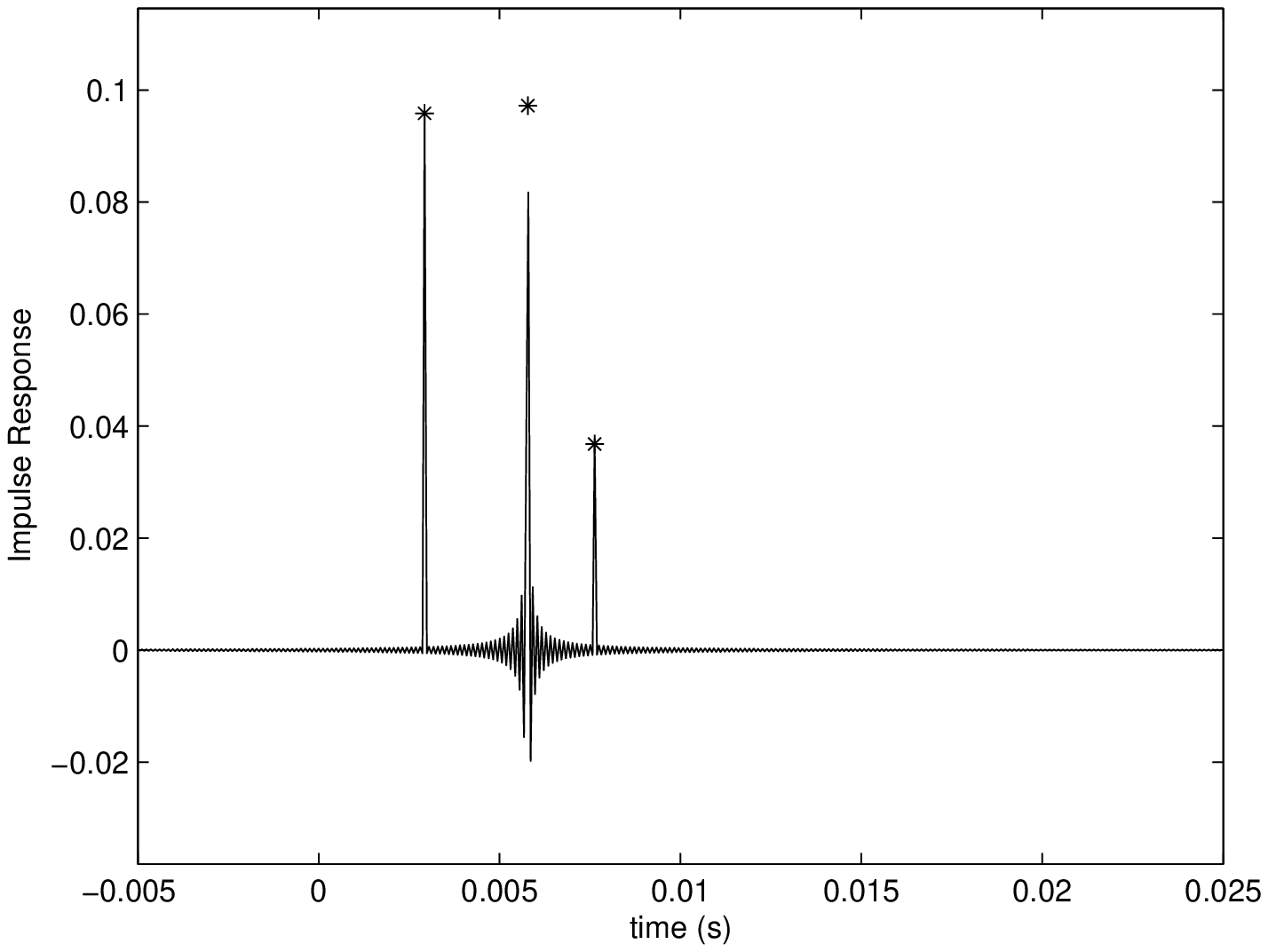}
\vspace*{5cm}
\caption{}
\label{fig:second_path_time}
\end{figure}

Figure \ref{fig:second_path_time}: Impulse Response of the secondary path,
* continuous time response, --- discrete-time model.

\begin{figure}[h]
\vspace*{2cm}
\centering
\includegraphics[width=10cm]{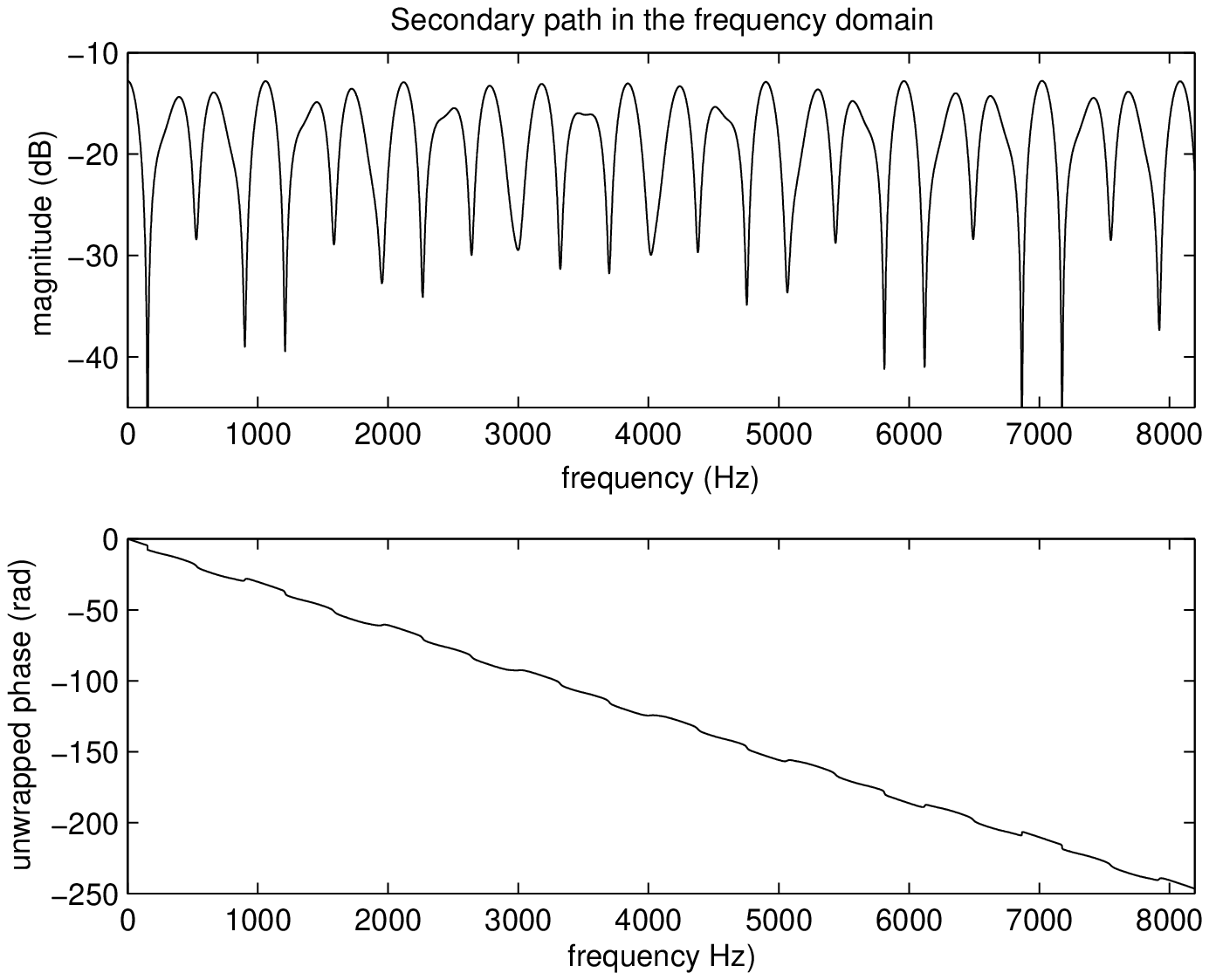}
\vspace*{5cm}
\caption{}
\label{fig:second_path_freq}
\end{figure}

Figure \ref{fig:second_path_freq}: Frequency Response Function of the secondary path.

\begin{figure}[ht]
\vspace*{3cm}
\centering
\includegraphics[width=10cm]{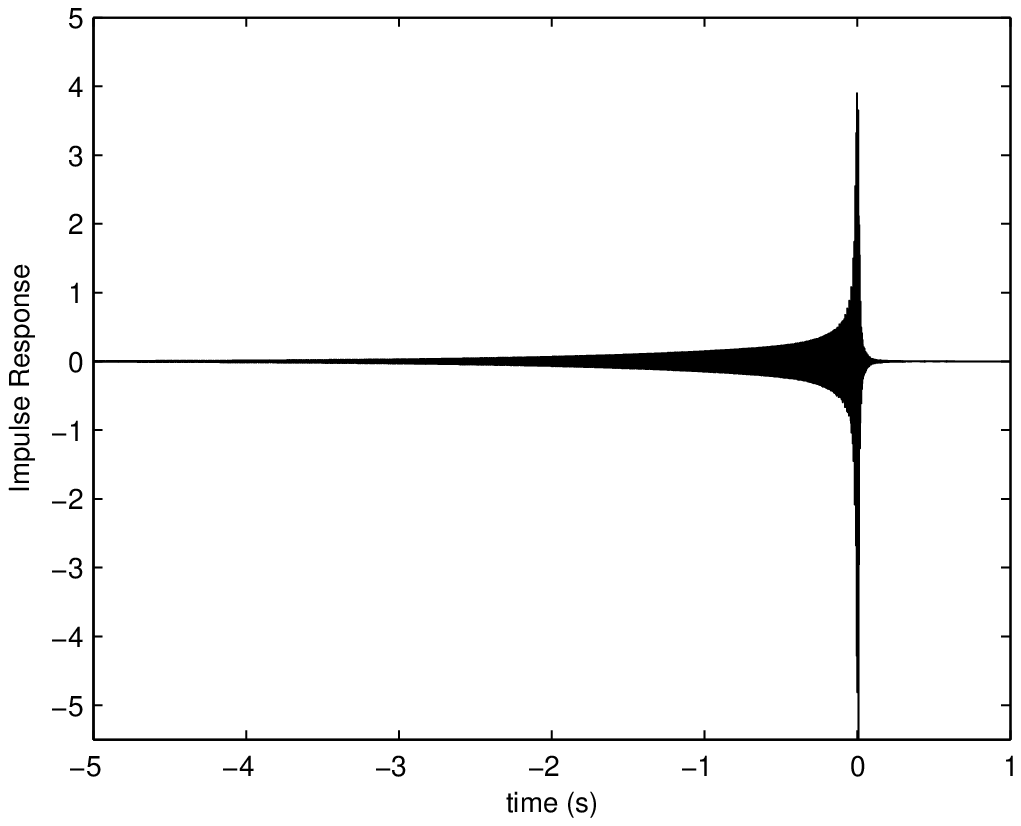}
\vspace*{5cm}
\caption{}
\label{fig:opt_temp}
\end{figure}

Figure \ref{fig:opt_temp}: Impulse Response of the optimal non-causal control filter.

\begin{figure}[ht]
\vspace*{3cm}
\centering
\includegraphics[width=10cm]{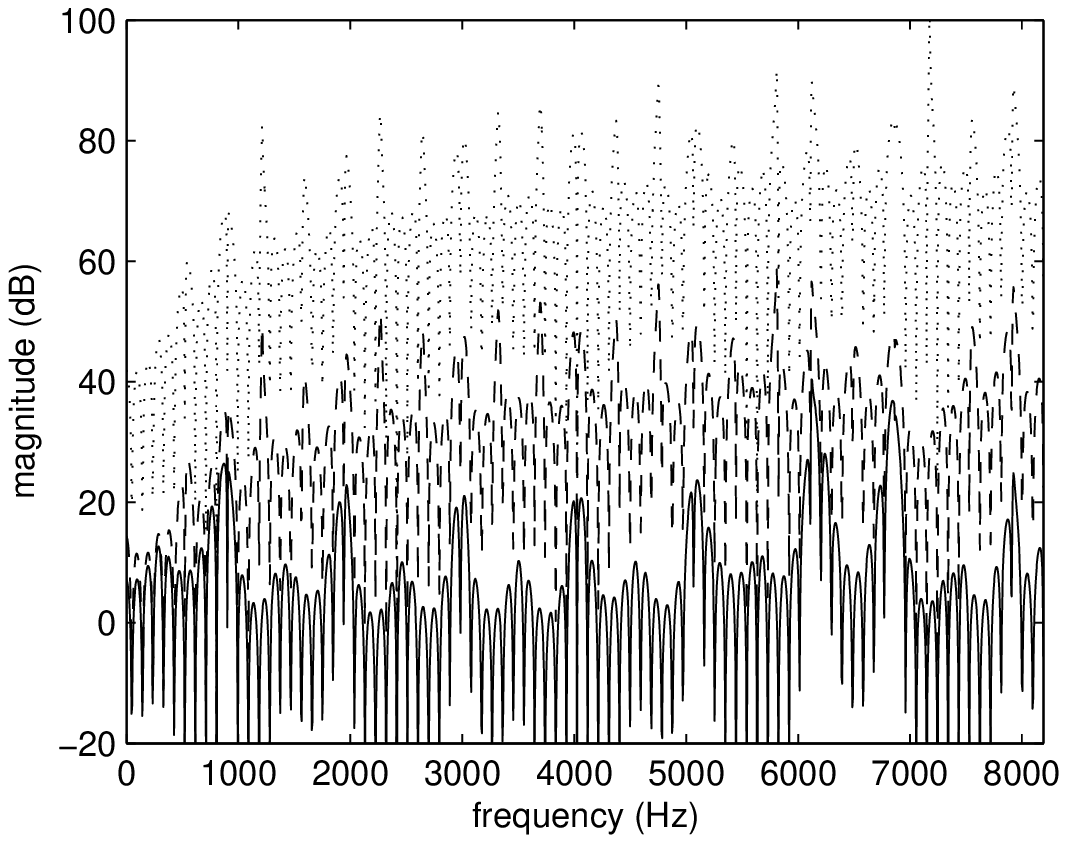}
\vspace*{5cm}
\caption{}
\label{fig:atten_tronq}
\end{figure}

Figure \ref{fig:atten_tronq}: Noise attenuation with causal truncation of optimal control filters,
--- without additionnal advance in the reference signal, - - with a 0.4s advance,
... with a 5s advance.

\begin{figure}[ht]
\vspace*{3cm}
\centering
\includegraphics[width=10cm]{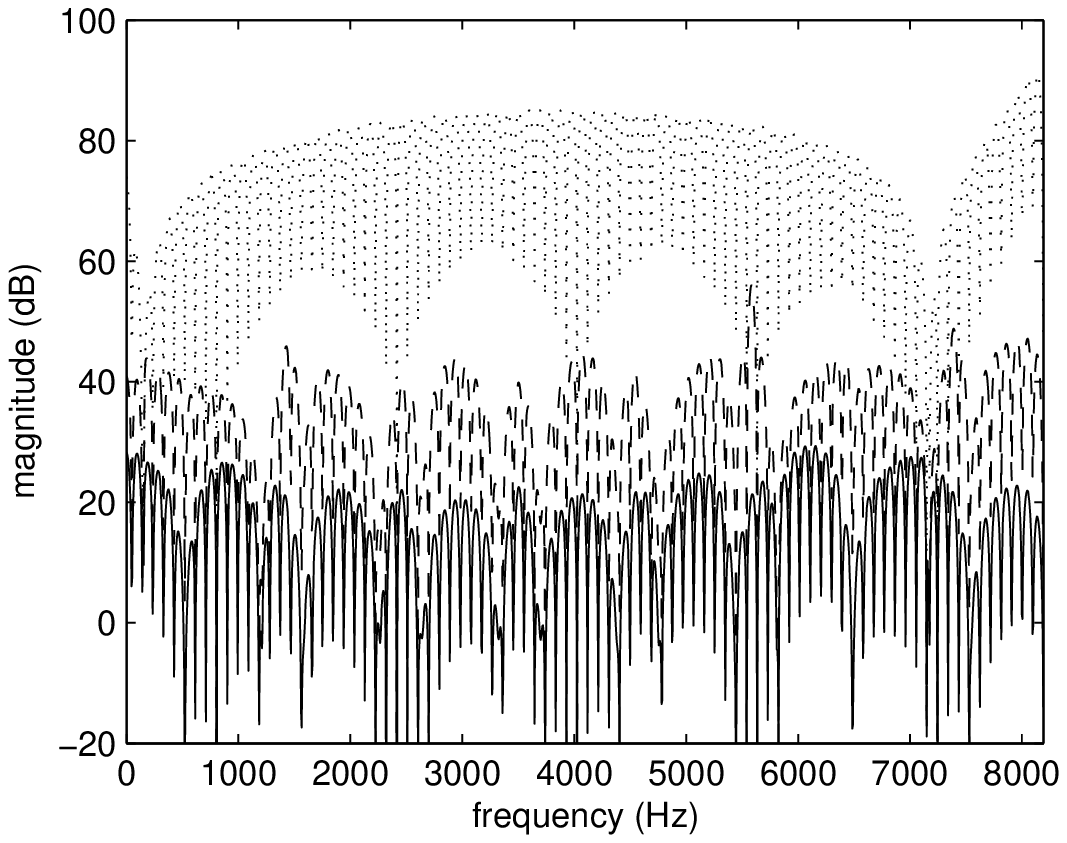}
\vspace*{5cm}
\caption{}
\label{fig:atten_opt_caus}
\end{figure}

Figure \ref{fig:atten_opt_caus}: Noise attenuation with causally-constrained optimal control filters
computed in the frequency domain, --- without additional advance in the reference signal,
- - with a 31.25ms advance, ... with a 400ms advance.

\begin{figure}[ht]
\vspace*{3cm}
\centering
\includegraphics[width=10cm]{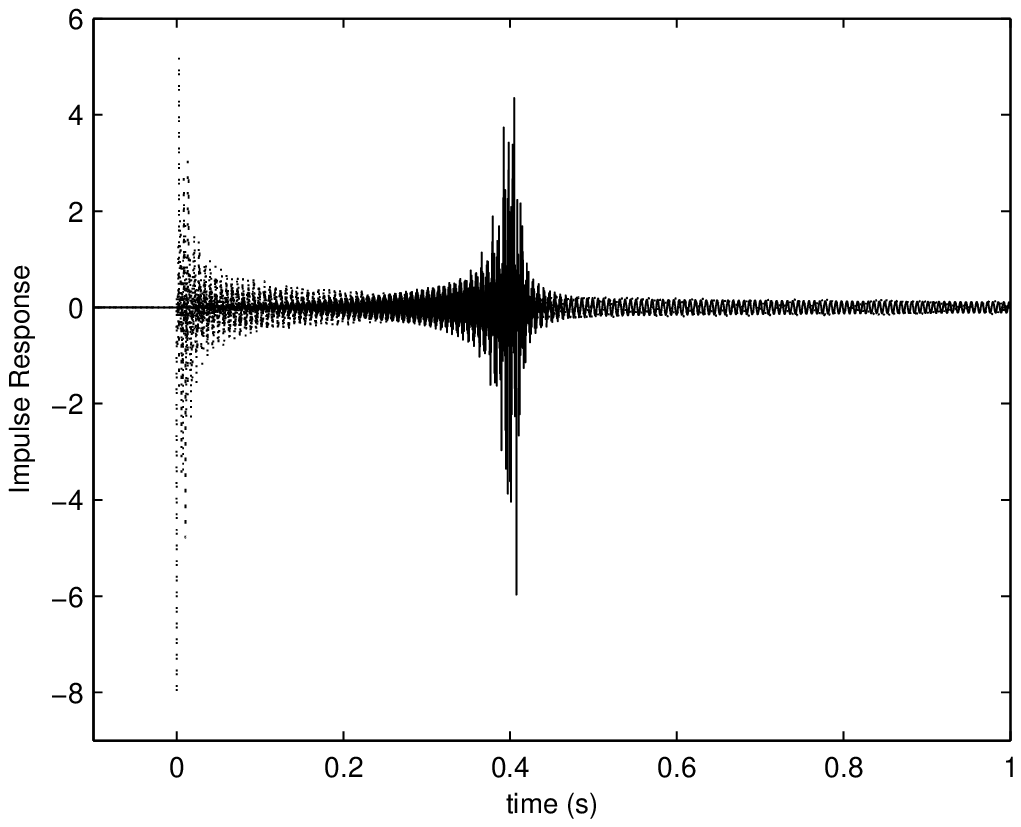}
\vspace*{5cm}
\caption{}
\label{fig:opt_caus_temp}
\end{figure}

Figure \ref{fig:opt_caus_temp}: Impulse Response of causally-constrained optimal control filters
computed in the frequency domain,  ... without additional advance in the reference signal,
--- with a 0.4s advance.

\begin{figure}[ht]
\vspace*{3cm}
\centering
\includegraphics[width=10cm]{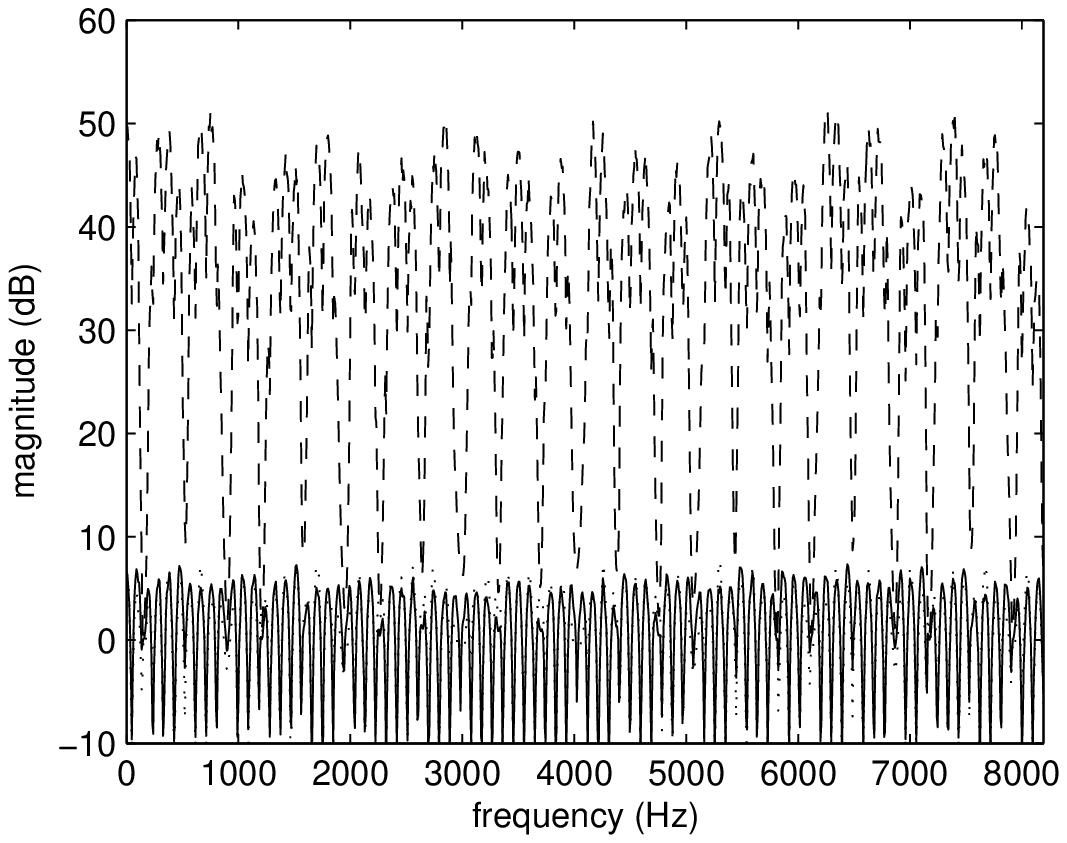}
\vspace*{5cm}
\caption{}
\label{fig:atten_fxlms_aver}
\end{figure}

Figure \ref{fig:atten_fxlms_aver}: Noise attenuation with a 1024-coefficient control FIR filter
updated with an FXLMS algorithm,
--- without additional advance in the reference signal, - - with a 31.25ms advance,
... with a 62.5ms advance.

\begin{figure}[ht]
\vspace*{3cm}
\centering
\includegraphics[width=10cm]{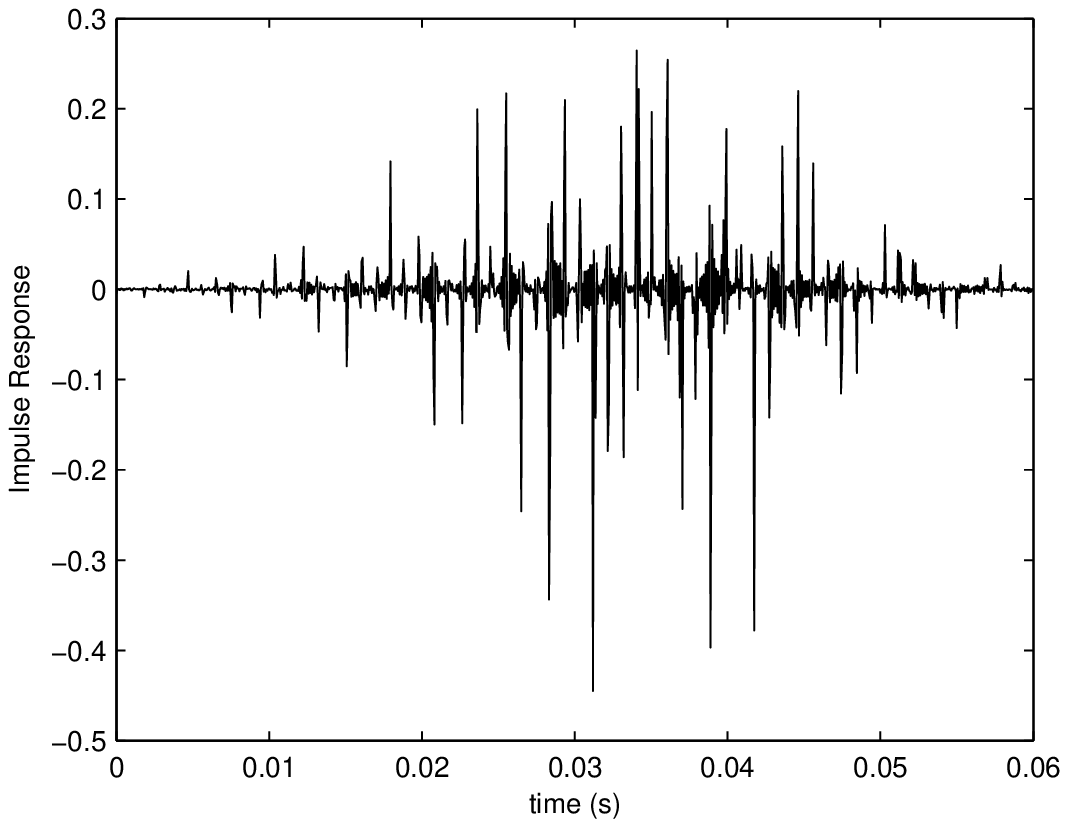}
\vspace*{5cm}
\caption{}
\label{fig:fxlms_temp}
\end{figure}

Figure \ref{fig:fxlms_temp}: Impulse response of a 1024-coefficient control FIR filter
frozen after minimization of the error signal, with a 31.25ms advance in the reference signal.

\begin{figure}[ht]
\vspace*{3cm}
\centering
\includegraphics[width=10cm]{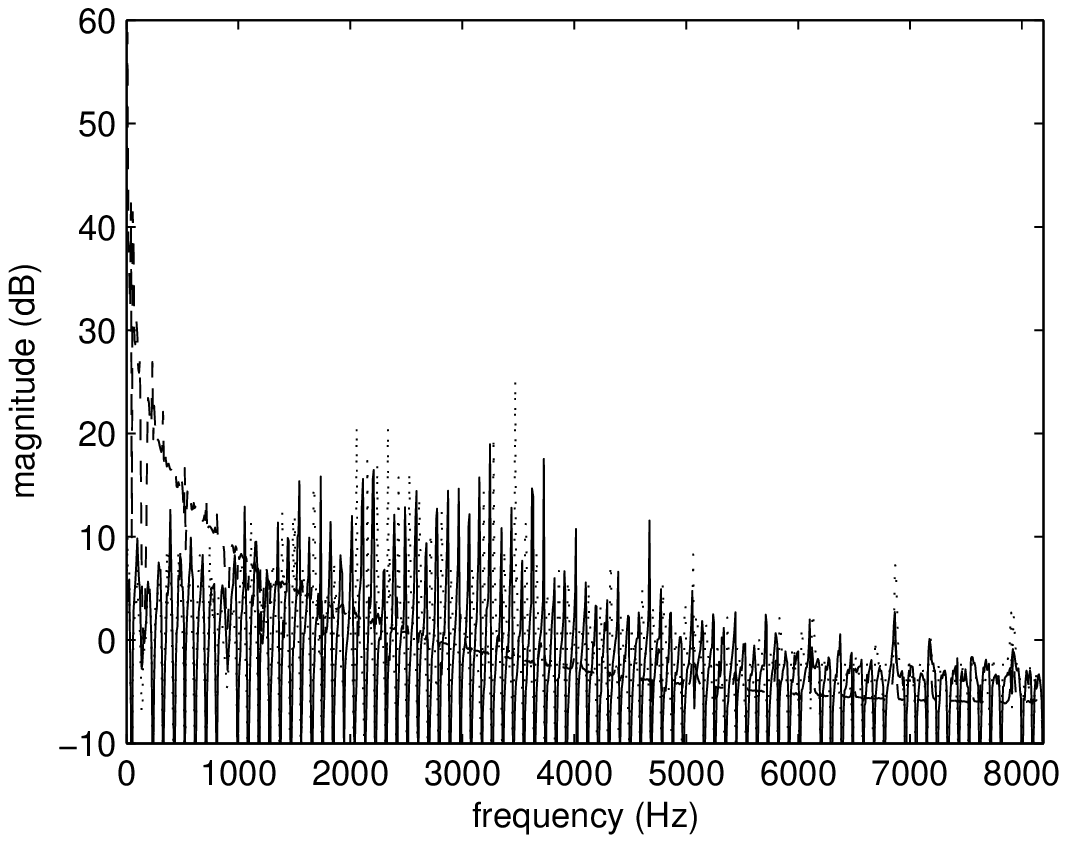}
\vspace*{5cm}
\caption{}
\label{fig:atten_fxlms_froz}
\end{figure}

Figure \ref{fig:atten_fxlms_froz}: Noise attenuation with 1024-coefficient control FIR filters
frozen after minimization of the error signal,
--- without additional advance in the reference signal, - - with a 31.25ms advance,
... with a 62.5ms advance.

\begin{figure}[ht]
\vspace*{3cm}
\centering
\includegraphics[width=10cm]{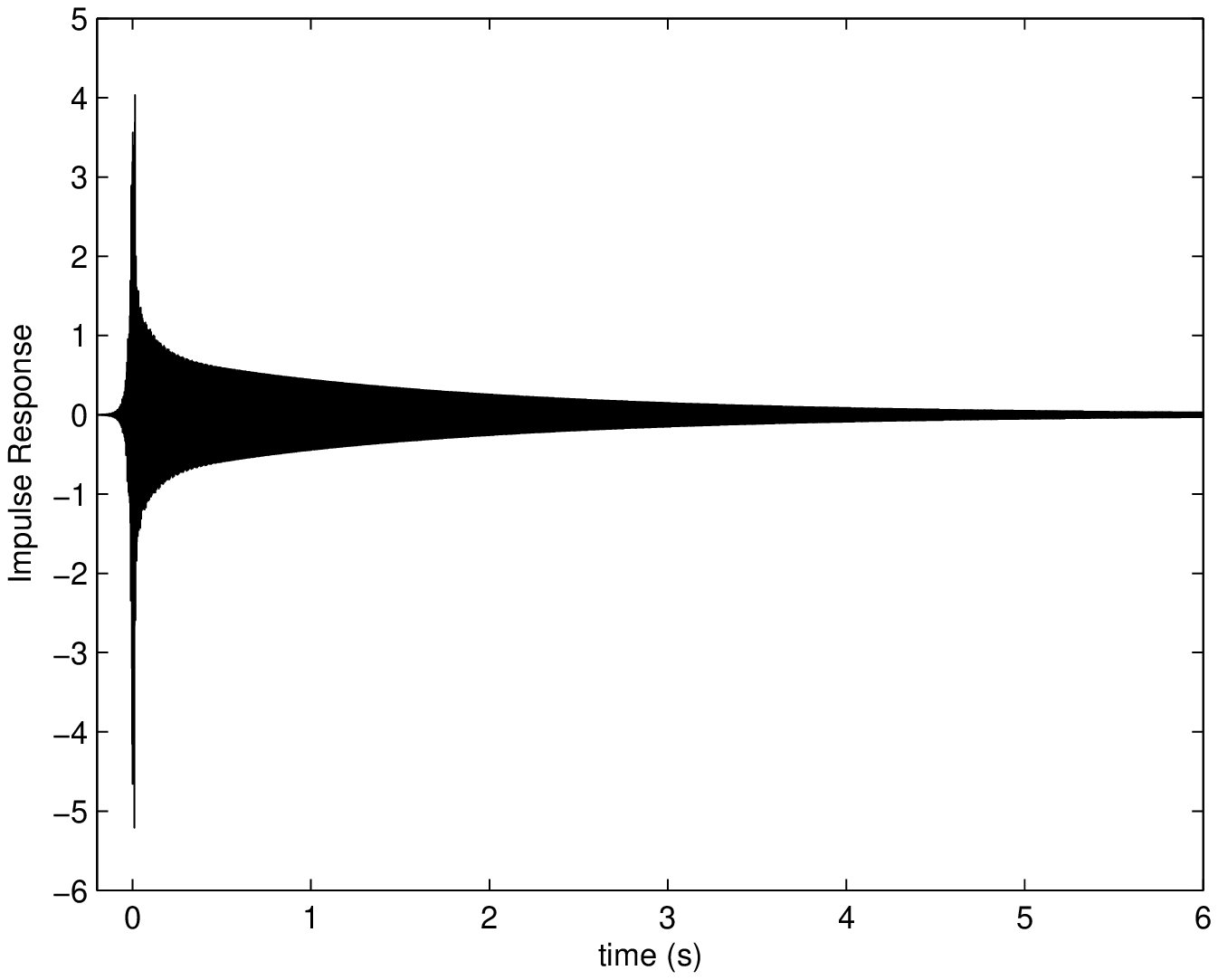}
\vspace*{5cm}
\caption{}
\label{fig:coin_moins_loin}
\end{figure}

Figure \ref{fig:coin_moins_loin}: Impulse Response of the optimal non-causal control filter for L=0.95m.

\begin{figure}[ht]
\vspace*{3cm}
\centering
\includegraphics[width=10cm]{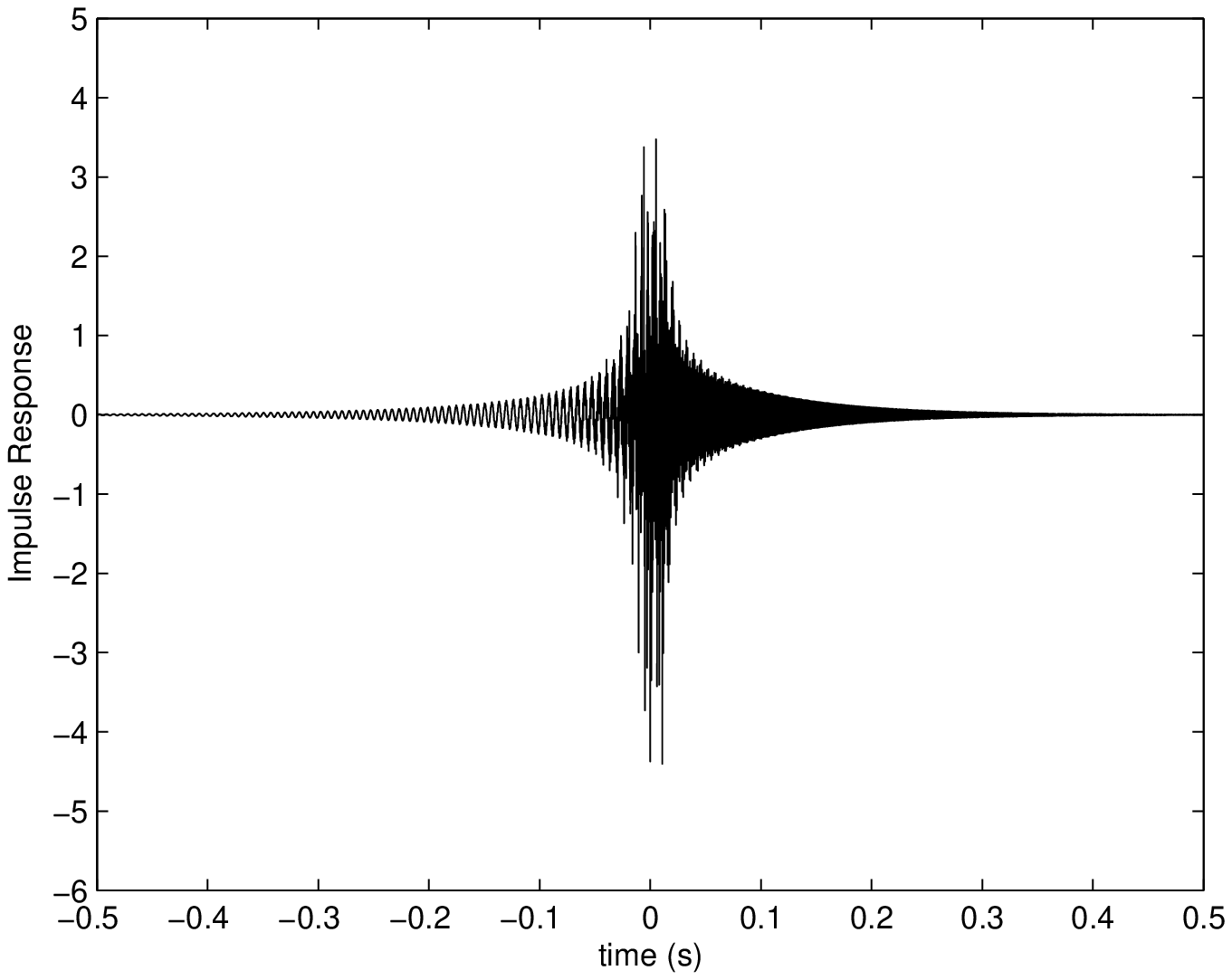}
\vspace*{5cm}
\caption{}
\label{fig:coin_plus_loin}
\end{figure}

Figure \ref{fig:coin_plus_loin}: Impulse Response of the optimal non-causal control filter for L=1.05m.

\begin{figure}[ht]
\vspace*{3cm}
\centering
\includegraphics[width=12cm]{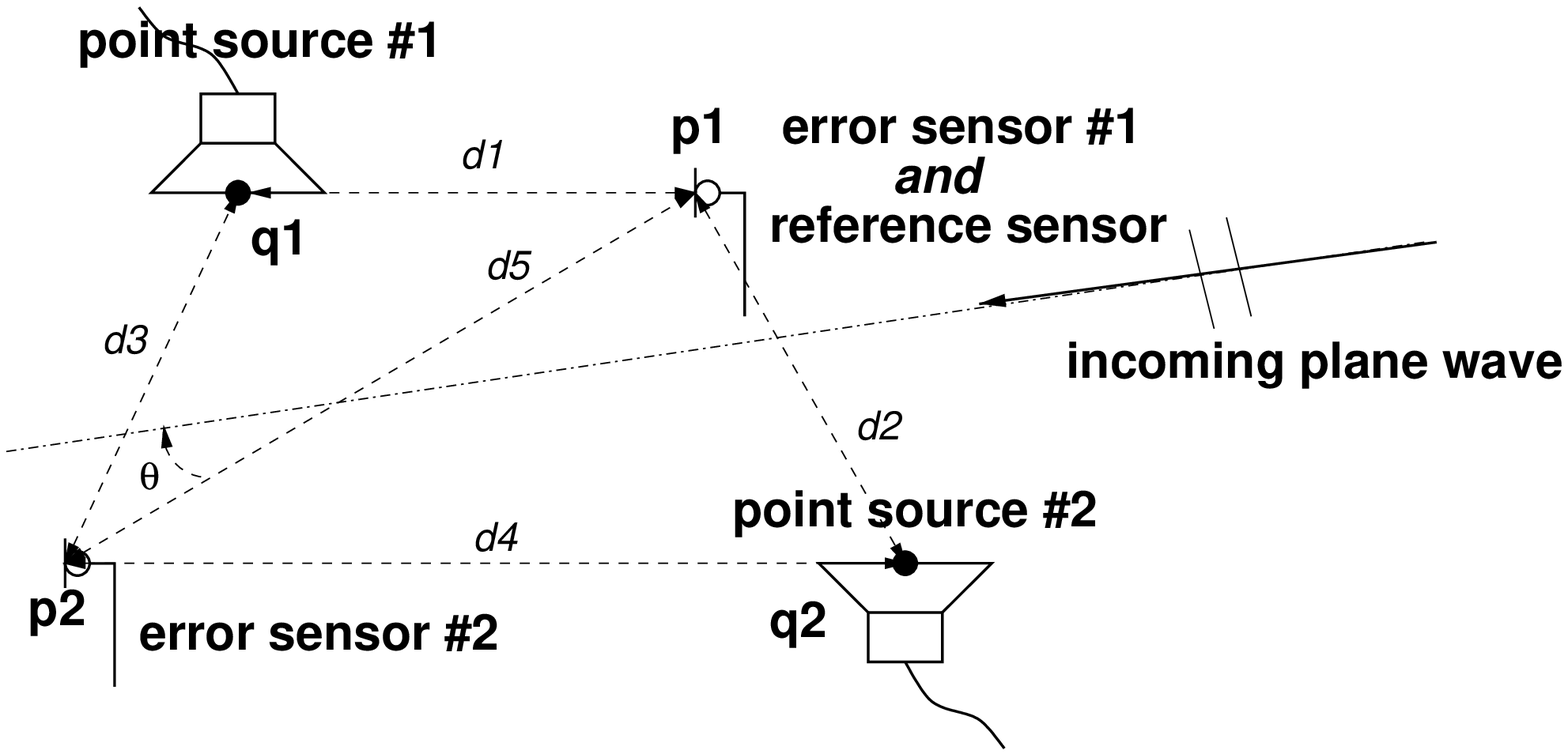}
\vspace*{5cm}
\caption{}
\label{fig:2-2set-upJSV}
\end{figure}

Figure \ref{fig:2-2set-upJSV}: A 2x2 ANC set-up in free field.

\begin{figure}[ht]
\vspace*{3cm}
\centering
\includegraphics[width=12cm]{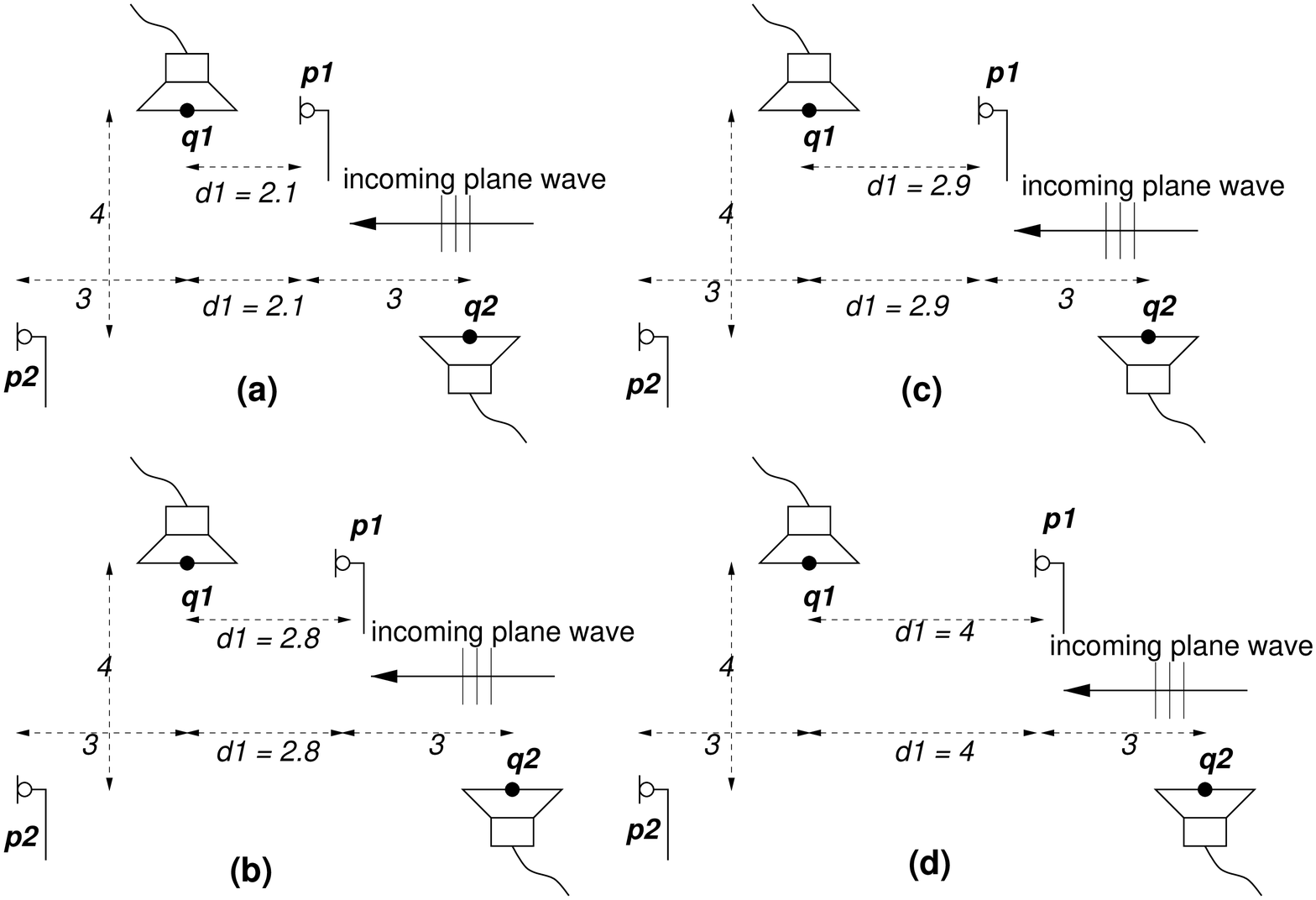}
\vspace*{5cm}
\caption{}
\label{fig:4cases}
\end{figure}

Figure \ref{fig:4cases}: 4 ANC arrangements leading to optimal filters very different with respect to causality.

\begin{figure}[ht]
\vspace*{3cm}
\centering
\includegraphics[height=5cm]{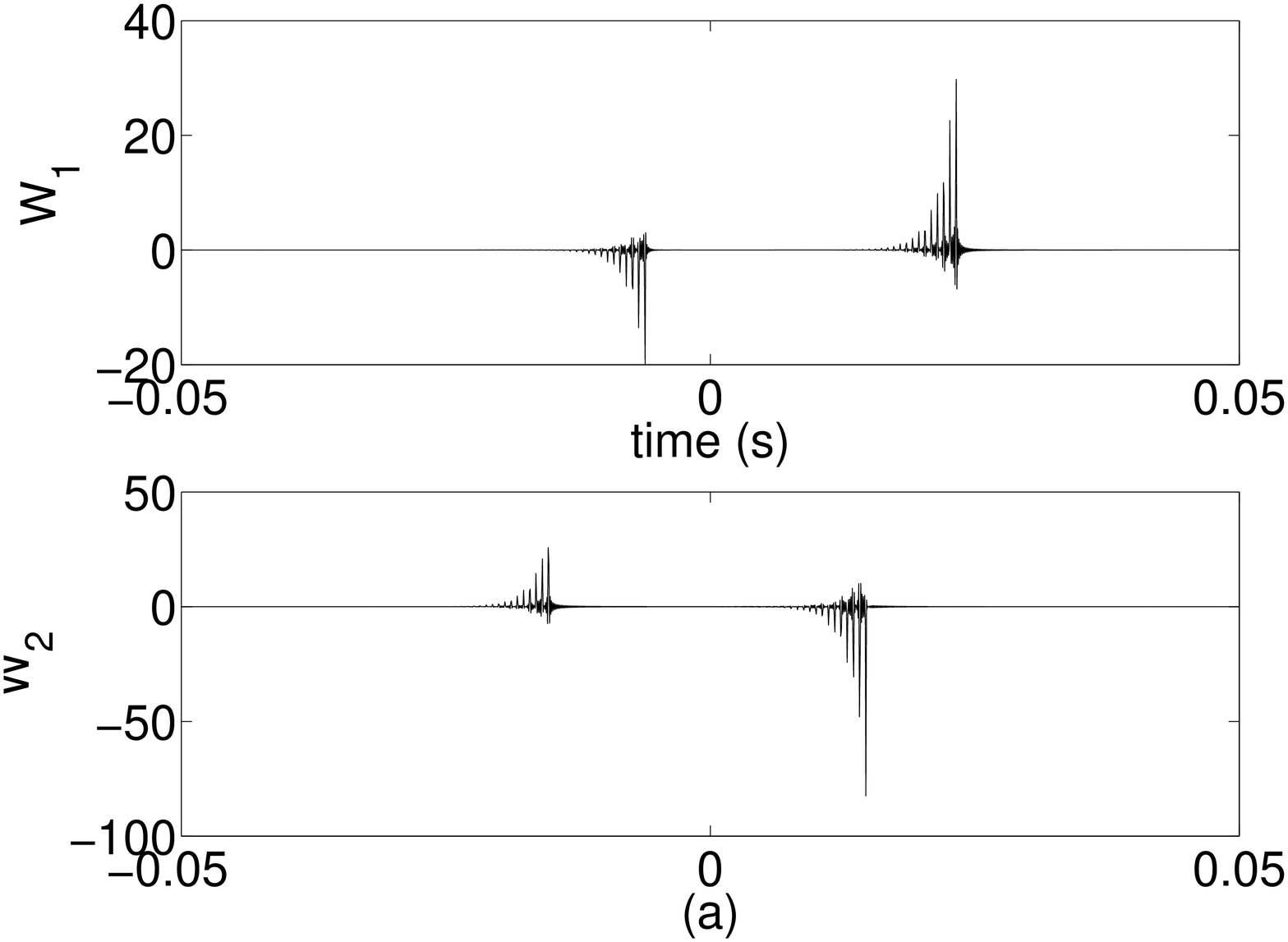}
\includegraphics[height=5cm]{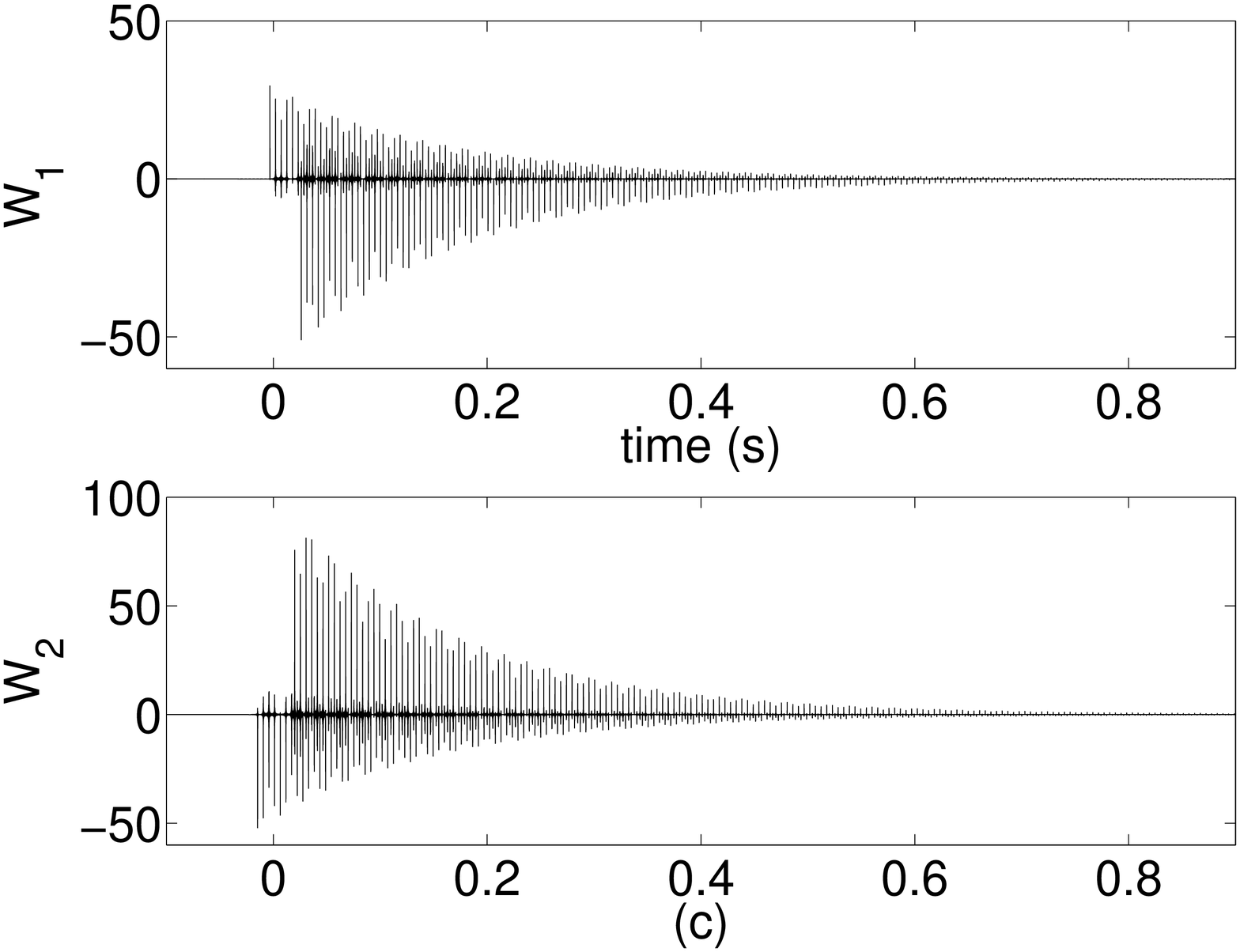}\\
\includegraphics[height=5cm]{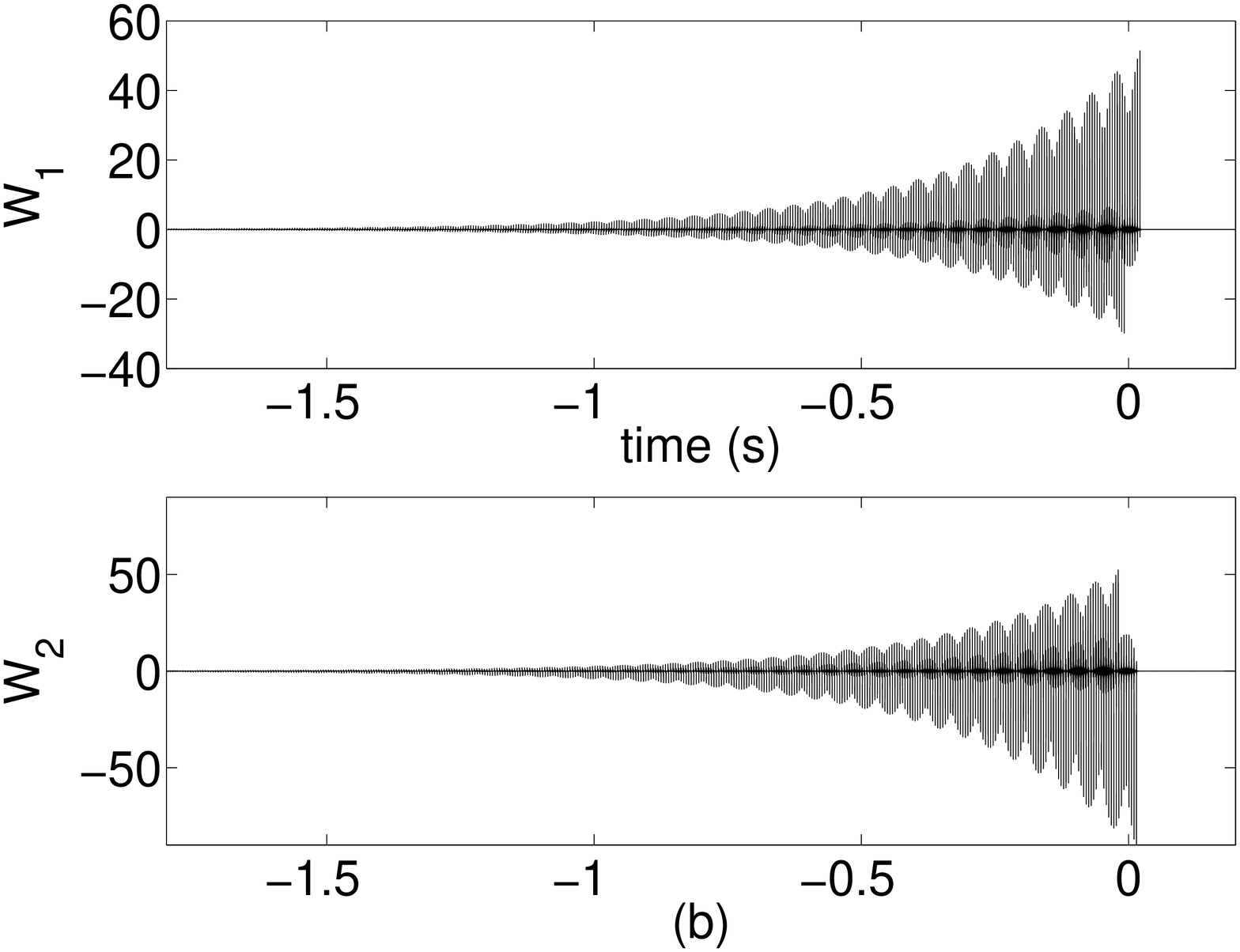}
\includegraphics[height=5cm]{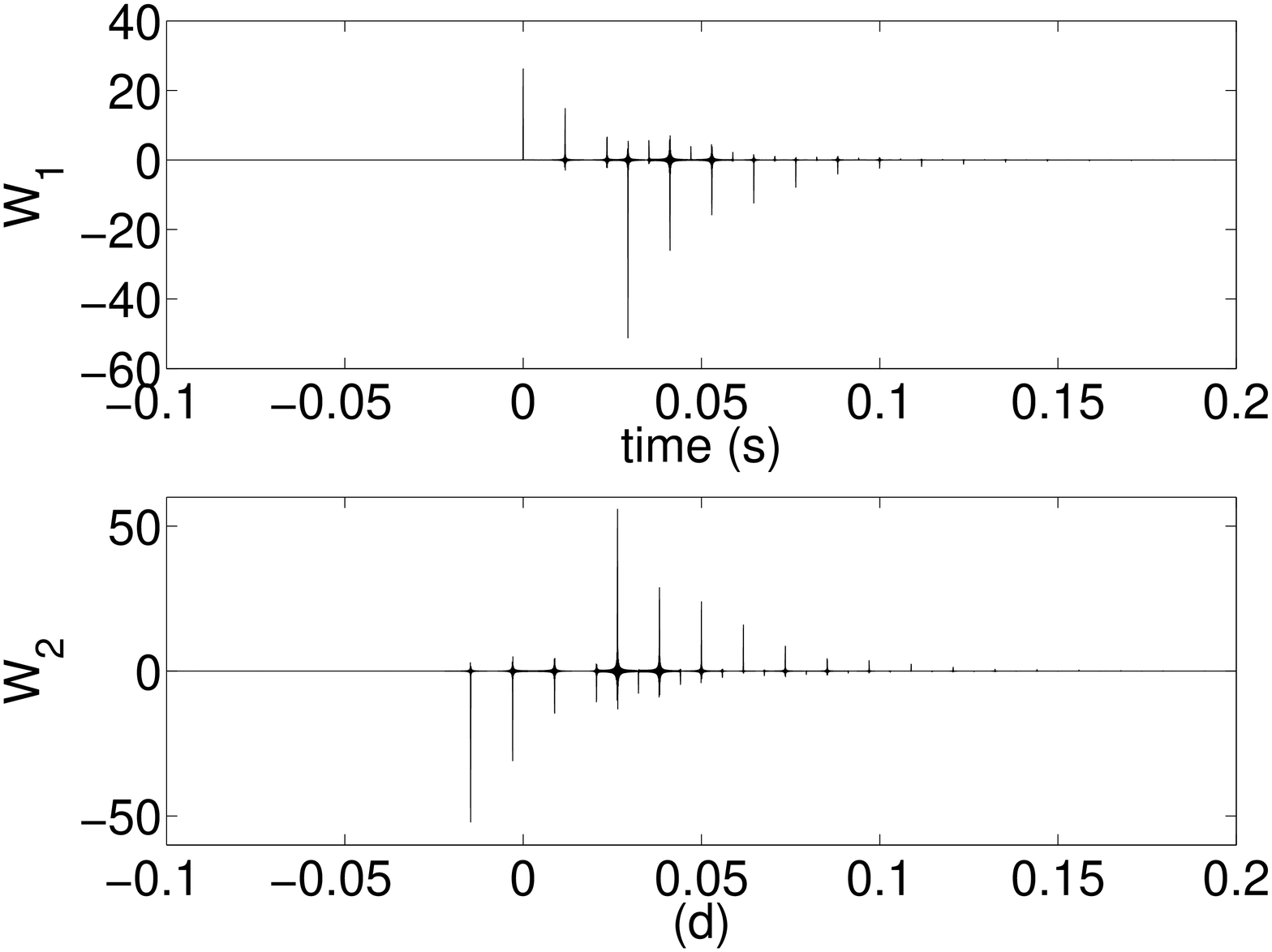}
\vspace*{5cm}
\caption{}
\label{fig:rep4cases}
\end{figure}

Figure \ref{fig:rep4cases}: Impulse Responses of the optimal control filters for the 4 arrangements shown
in figure~\ref{fig:4cases}.

\begin{figure}[ht]
\vspace*{3cm}
\centering
\includegraphics[width=11cm]{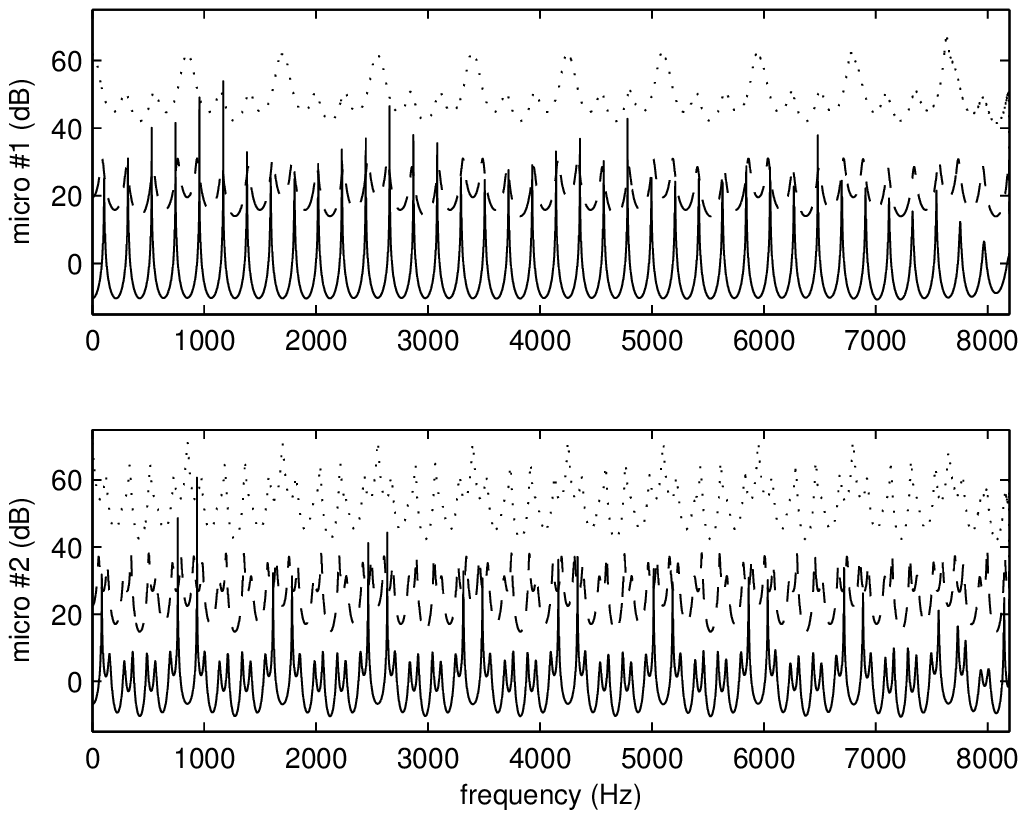}
\vspace*{5cm}
\caption{}
\label{fig:atten_tronq2x2}
\end{figure}

Figure \ref{fig:atten_tronq2x2}: Noise attenuation with causal truncation of optimal control filters,
--- with an 8.24ms additional time advance in the reference signal, - - with a 1s advance,
... with a 2s advance.

\begin{figure}[ht]
\vspace*{3cm}
\centering
\includegraphics[width=10cm]{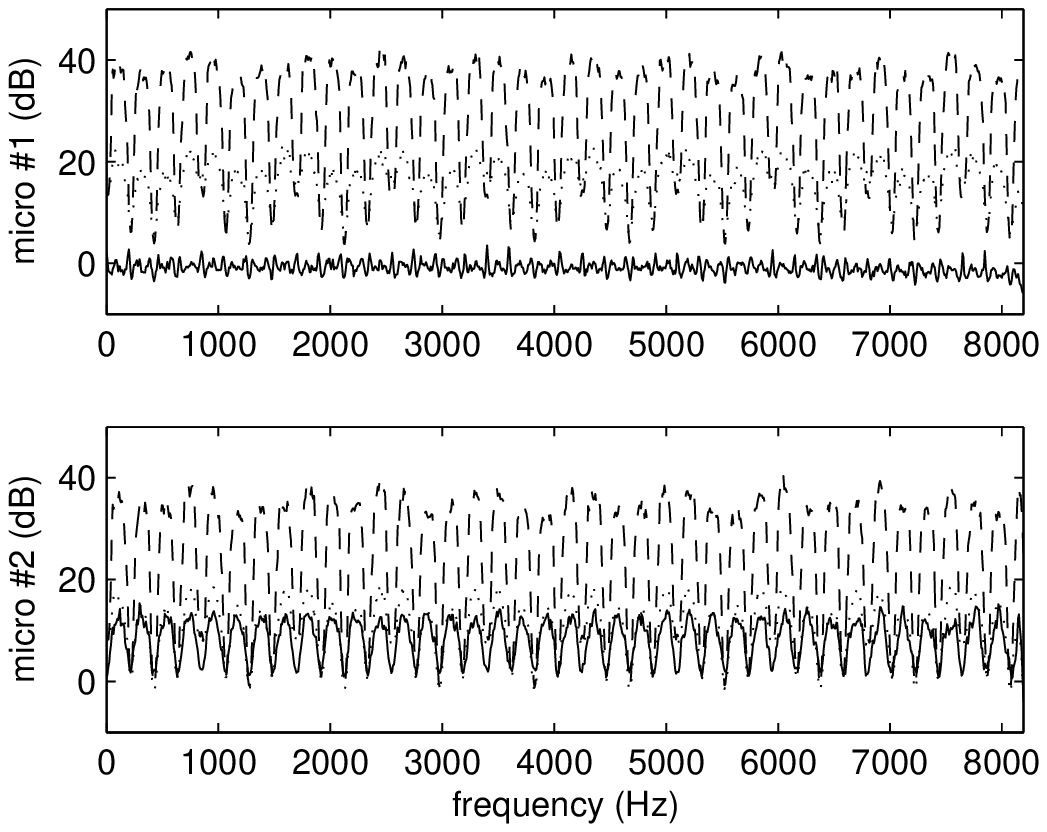}
\vspace*{5cm}
\caption{}
\label{fig:atten_fxlms_aver2x2}
\end{figure}

Figure \ref{fig:atten_fxlms_aver2x2}: Noise attenuation with a 1024-coefficient control FIR filter
updated with an FXLMS algorithm,
--- with an 8.24ms additional time advance in the reference signal, - - with a 31.25ms advance,
... with a 62.5ms advance.

\begin{figure}[ht]
\vspace*{3cm}
\centering
\includegraphics[width=10cm]{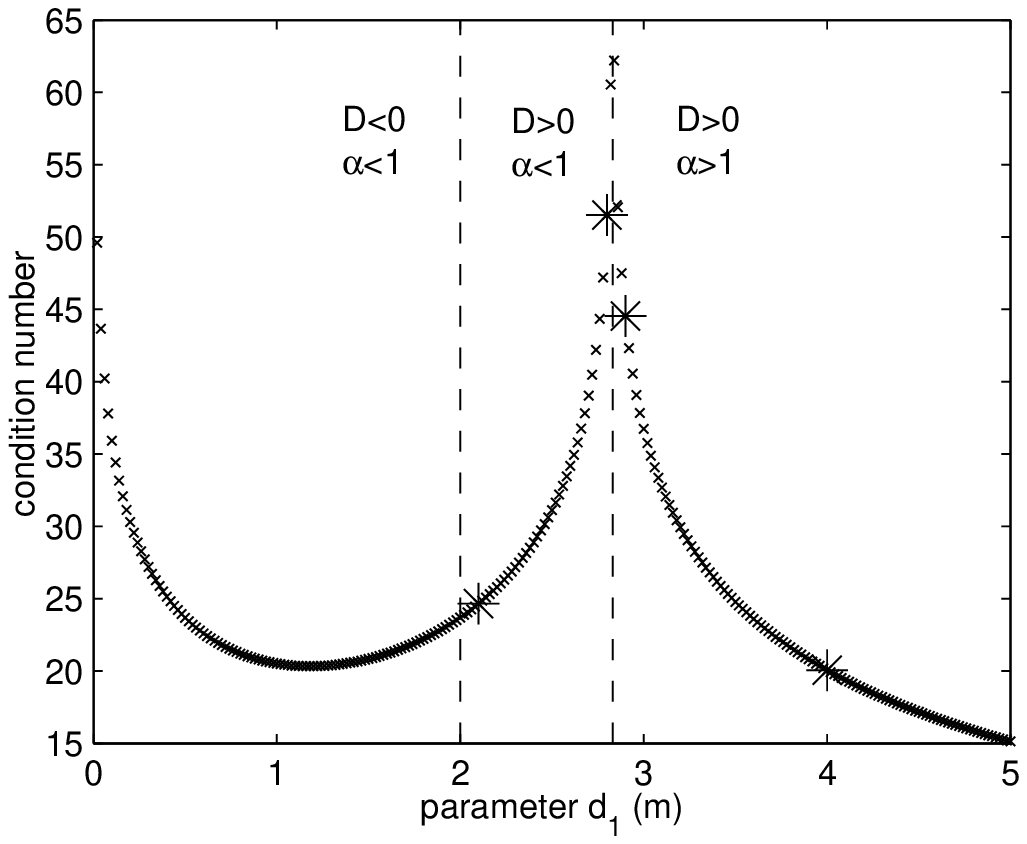}
\vspace*{5cm}
\caption{}
\label{fig:conditionnement}
\end{figure}

Figure \ref{fig:conditionnement}: Condition number of the secondary path matrix at low-frequency as a function of 
geometric parameter $d_1$, for $d_2=d_3=5$m and $d_4=d_1+6$m.
The stars mark out the values corresponding to the arrangements diplayed in figure \ref{fig:4cases}. 

\newpage

\newpage

\newpage

\newpage

\newpage

\newpage

\newpage

\newpage

\newpage

\newpage

\newpage

\newpage

\newpage

\newpage

\newpage

\newpage

\newpage

\newpage

\newpage

\end{document}